\PassOptionsToPackage{table}{xcolor}
\documentclass[sigconf,screen]{acmart}

\copyrightyear{2026}
\acmYear{2026}
\setcopyright{cc}
\setcctype{by}
\acmConference[PACT '26]{International Conference on Parallel Architectures and Compilation Techniques}{October 19--22, 2026}{Chicago, IL, USA}
\acmBooktitle{International Conference on Parallel Architectures and Compilation Techniques (PACT '26), October 19--22, 2026, Chicago, IL, USA}
\acmDOI{10.1145/3838684.3846860}
\acmISBN{979-8-4007-2915-7/2026/10}


\usepackage{multirow}
\usepackage{makecell}
\usepackage{siunitx}
\usepackage{subcaption}
\usepackage{pifont}
\usepackage{enumitem}
\usepackage{listings}
\usepackage{stfloats}
\usepackage[ruled,linesnumbered,vlined]{algorithm2e}
\usepackage{float}
\newcounter{matrixctr}
\renewcommand{\thematrixctr}{M\arabic{matrixctr}}

\definecolor{acmblue}{RGB}{0, 0, 255}    
\definecolor{acmpink}{RGB}{215, 30, 115}  
\definecolor{acmpurple}{RGB}{120, 0, 200} 
\definecolor{acmgreen}{RGB}{0, 128, 0}    
\definecolor{acmred}{RGB}{200, 0, 0}      

\lstdefinelanguage{MLIR}{
  morecomment=[l]{//},
  morestring=[b]",
  alsoletter={., ?, x, <, >, ^, _},      
  sensitive=true,
  morekeywords={
    module, func.func, return, arith.constant, arith.mulf, arith.addf, memref.load, vector.load,
    linalg.generic, linalg.fill, linalg.yield, tensor.empty,
    affine_map, sparse_tensor.encoding, vector.broadcast, arith.cmpi, arith.extui, vector.reduction, arith.index_cast, arith.addi, arith.select, vector.from_elements,
    map, dense, compressed, indexing_maps, iterator_types, ins, outs
  },
  morekeywords=[2]{index, f64, tensor, xf64, ?, memref, <?xf64>, int, double},
  morekeywords=[3]{scf.for, scf.while, scf.if, scf.yield, scf.condition, else, while, for, if, ?, \&\&, do, ^bb0},
  keywordstyle=\color{acmblue}\bfseries,
  keywordstyle=[2]\color{acmpurple},
  keywordstyle=[3]\color{acmpink}\bfseries,
  commentstyle=\color{acmgreen}\itshape,
  stringstyle=\color{acmred},
}

\lstdefinestyle{acm_mlir_style}{
  language=MLIR,
  basicstyle=\ttfamily\small,   
  upquote=true,
  aboveskip={1\baselineskip},
  belowskip={1\baselineskip},
  columns=fullflexible,
  framesep=4pt,
  breaklines=true,                   
  breakatwhitespace=true,            
  tabsize=2,
  numbers=left,
  numberstyle=\tiny\color{gray!90},
  numbersep=8pt,
  xleftmargin=1.5em,                 
}

\newcommand{\confnum}[1]{{\setlength{\fboxsep}{1.5pt}\fbox{\small #1}}}

\newcommand{\eg}{\textit{e.g.,}~}
\newcommand{\ie}{\textit{i.e.,}~}

\newcommand{\scfwhile}{\texttt{scf.while}~}

\newcommand{\sparsedialect}{\texttt{SparseTensor}~}

\newcommand{\vectordialect}{\texttt{Vector}~}

\newcommand{\name}{\textsc{Splyce}\xspace}

\begin{document}

\title{\name: SIMD Vectorization of Sparse Coiteration}

\author{Kabilan Mahathevan}
\orcid{0009-0003-4980-4886}
\affiliation{%
  \institution{Virginia Tech}
  \city{Blacksburg}
  \country{USA}
}
\email{kabilan@vt.edu}

\author{Poorna Gunathilaka}
\orcid{0009-0005-7895-4722}
\affiliation{%
  \institution{Virginia Tech}
  \city{Blacksburg}
  \country{USA}
}
\email{poornag@vt.edu}

\author{Kirshanthan Sundararajah}
\orcid{0000-0001-6384-062X}
\affiliation{%
  \institution{Virginia Tech}
  \city{Blacksburg}
  \country{USA}
}
\email{kirshanthans@vt.edu}


\begin{abstract}
Sparse tensor contractions are bottlenecked by sparse-sparse coiteration loops that resist standard loop vectorization. 
We present \name, an auto-vectorization framework in MLIR that overcomes this through a dual-path execution model. 
By decoupling coordinate intersection from pointer management via selective predication, \name inherently eliminates data-dependent branches as a side effect, allowing modern superscalar engines to maximize instruction-level parallelism and hide memory latency. 
Beyond simple branch elimination, our transformation exposes independent computation that can be executed concurrently, increasing functional-unit utilization that would otherwise be constrained by sequential dependencies. 
Evaluation across foundational sparse tensor kernels demonstrates performance ranging from 1.96$\times$ to 2.86$\times$ on synthetic inputs, with consistent speedups sustained across a vast majority of irregular real-world datasets from the SuiteSparse collection. 
Ultimately, \name demonstrates that by converting unpredictable control-flow into a predictable data stream, compiler-driven speculation can effectively reconcile the memory efficiency of compressed storage with the execution-unit throughput of modern superscalar architectures.
\end{abstract}

\maketitle

\section{Introduction} \label{sec:introduction}

Dense tensor optimizations are well-established~\cite{distal,dastac,Li2019Analytical,Senanayake2020A,distal,structured,Chen2018Learning}, but the proliferation of sparse data structures in modern workloads has shifted focus toward formats that compress data to save memory and reduce compute bounds~\cite{Qin2021Extending, Smith2015Tensor-matrix, taco, Willcock2006Accelerating, Chen2019Performance-Aware, Chou2018Format}.
However, the complex memory layouts required for sparse compression complicate iteration and automated code generation, especially when compared to the straightforward, random-access nature of dense arrays. 
To address this, tensor compilers increasingly rely on the Compressed Sparse Fiber (CSF) format for its generalizability~\cite{Smith2015Tensor-matrix}. 
By modeling a tensor as a hierarchical tree, CSF allows each dimension to be annotated independently as dense, compressed, or singleton. 
Compressed dimensions explicitly track non-zero elements using position and coordinate arrays, while singleton levels represent unique coordinate entries without the overhead of pointers. 
Dense dimensions rely on implicit, fixed-size indexing, culminating in a unified contiguous array of non-zero values at the leaf level.

To navigate the complexities of generating code for diverse sparse data layouts, advanced tensor compilers have been developed. 
Most notably, building upon the pioneering success of TACO~\cite{taco}, the MLIR framework integrated a dedicated sparse tensor dialect~\cite{mlir_sparse} that adapts TACO's principles for a modern, multi-level compilation ecosystem. 
Consequently, active research is heavily focused on designing and implementing novel optimization strategies across the various abstraction layers within this infrastructure.

\begin{figure}[t]
    \centering
    \includegraphics[width=0.65\columnwidth]{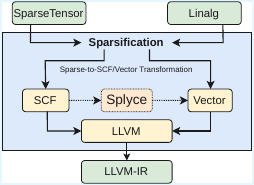}
    \caption{\sparsedialect dialect lowering with \name.}
    \label{fig:lowering_path}
    \vspace{-1em}
\end{figure}

The \texttt{SparseTensor} dialect operates as a compilation infrastructure that heavily leverages the progressive lowering ecosystem of MLIR, seamlessly integrating with the \texttt{Linalg}, \texttt{SCF}, and \texttt{Vector} dialects. 
Within this framework, core computations are abstractly defined using \texttt{linalg.generic} operations coupled with per-tensor sparse encoding annotations. 
As illustrated in Figure~\ref{fig:lowering_path}, the standard code generation pipeline progressively lowers these abstractions to achieve high performance, actively targeting the \texttt{Vector} dialect to exploit SIMD hardware whenever the underlying data layout permits. 
However, a major caveat of sparse tensor compilation is that the execution efficiency is inherently bound to the distribution of non-zeros in input data---a dynamic characteristic that the static compile-time optimizations largely fail to exploit~\cite{vecRC}.

A critical example of this limitation manifests in the \textit{two-finger} \texttt{scf.while} loop structures (\ie sparse coiteration loop) generated during the lowering process. 
This coiteration loop is a fundamental requirement of sparse tensor computations when a shared or contracting dimension is compressed on both inputs. 
To traverse this contracting dimension, the generated code must simultaneously advance two cursors through the coordinate arrays of the respective tensors that is similar to performing set intersection operation~\cite{lemire2015decoding,boncz2005monetdb,graefe1993query,smith1981identification,zhao2022polyhedral}. 
Crucially, for intersection operations, the inner computation is strictly predicated on a coordinate match; it executes exclusively when both cursors point to the same index. 
Because these pointers must advance dynamically based on runtime coordinate comparisons, this data-dependent control-flow inherently breaks standard loop optimizations, most notably vectorization~\cite{taco, Chou2018Format}.

The introduction of divergent \texttt{if-else} control-flow within the \texttt{scf.while} loop and the data-dependent loop bounds structurally prevents standard SIMD auto vectorization. 
While compilers can circumvent this by  evaluating the coiteration space as a dense kernel to eliminate branching, the resulting computational overhead negates the memory and performance benefits of the sparse representation overwhelmingly.

In this paper, we demonstrate that the throughput advantages of vectorized execution can actually outweigh the penalty of processing a calculated fraction of structural zeros. 
To achieve this, we introduce \textbf{\name}, a novel MLIR optimization pass that intercepts the sparsification pipeline between the \texttt{SCF} and \texttt{Vector} dialects (Figure~\ref{fig:lowering_path}). 
\name transforms the standard scalar sparse coiteration into a dual-path execution consisting of a vectorized fast-path and a scalar slow-path (Figure~\ref{fig:vectorized}).

The fast-path iterates at the hardware vector width, utilizing predicated execution and vector masks to safely perform speculative SIMD operations.  
This transformation inherently eliminates data-dependent branches, allowing modern superscalar engines to maximize instruction-level parallelism (ILP) by overlapping coordinate matching with core tensor arithmetic. Specifically, the contributions of this work are as follows:
 
\begin{itemize}[noitemsep, topsep=0pt, parsep=0pt, partopsep=0pt]

    \item \textbf{\name Vectorization:} We implement the \name pass in MLIR to natively vectorize sparse coiterations via a branchless fast-path, enabling SIMD execution on compressed formats.
    
    \item \textbf{Analysis of Execution Phases:} We analyze sparse coiteration phases to identify the optimal balance between vector and scalar execution. We establish that a hybrid approach prevents execution port monopolization.
    
    \item \textbf{Performance Trade-off Analysis:} We quantify the cost-benefit of speculative vectorization, achieving a 2.38$\times$ geometric mean speedup across five kernels on synthetic datasets with 5\% sparsity. 
    
    \item \textbf{Sparsity and Parallel Scaling:} We demonstrate that \name maintains robust performance across tensor sparsities from 0.01\% to 10\% and real-world distributions, while achieving near-linear scalability across parallel multi-core environments.
\end{itemize}

To improve readability throughout the paper, we present code snippets in a simplified, hybrid MLIR/C-like form. 
Additionally, for brevity, we use the term \textit{coiteration} to refer specifically to \textit{sparse coiteration} (formally detailed in Section~\ref{subsec:sparsecoiteration}) throughout the remainder of this paper.
The remainder of this paper is structured as follows. 
Section~\ref{sec:background} provides essential background and Section~\ref{sec:approach} details the design of the \name approach. 
We illustrate the evaluation of our approach in Section~\ref{sec:evaluation}.
Section~\ref{sec:related_work} reviews related work, and we discuss future work in Section~\ref{sec:discussion}.
Finally, we conclude the paper in Section~\ref{sec:conclusion}.

\section{Background} 
\label{sec:background}

\subsection{Sparse Tensor Compilers (STCs)}
The proliferation of highly sparse tensors in modern applications, ranging from graph analytics to deep learning and quantum chemistry, has necessitated a fundamental shift in computational strategies for tensor contractions~\cite{kolda2009tensor, hirata2003tensor}.
Historically, the challenge of sparsity has been addressed by relying on manually crafted and tuned kernel libraries~\cite{Shi2016Tensor, intelMKL, cusparse, dongarra1990set}.
While these libraries deliver near-peak hardware utilization for basic linear algebra routines like SpGEMM and SpMV, they suffer from a severe scalability crisis.
Developing, optimizing, and maintaining hand-written kernels for every conceivable tensor expression across every permutation of compressed data formats, and targeting increasingly diverse hardware architectures, is fundamentally intractable~\cite{taco}. 

Consequently, modern frameworks have shifted toward compiler-driven generation~\cite{taco, ye2023sparsetir, mlir_sparse, ahrens2025finch}, successfully enabling the generalized traversal of compressed sparse tensors without materializing them into memory-intensive dense formats. 
The advent of the Tensor Algebra Compiler (TACO)~\cite{taco} marked a paradigm shift by introducing the iteration lattice---a mathematical framework that decouples the tensor expression from its underlying layout to automatically synthesize coiteration control-flow.

To achieve this generality, modern STCs rely on Compressed Sparse Fiber (CSF) abstractions. 
CSF models any sparse tensor as a hierarchical tree where each axis is defined independently, where each axis (or dimension) of a tensor is defined as either dense or compressed. 
By composing these level-wise annotations, an STC can natively represent and compile code for a vast spectrum of traditional data structures. 
For instance, a 2D matrix encoded with an outer-dense axis and an inner-compressed axis yields the standard Compressed Sparse Row (CSR) format. 
Reversing this to compressed-dense yields the Compressed Sparse Column (CSC) format, while a compressed-compressed annotation generates the Doubly Compressed Sparse Row (DCSR) format~\cite{Chou2018Format}.

\subsection{MLIR \texttt{SparseTensor} Dialect}

Building upon these generalized STC frameworks, the Multi-Level Intermediate Representation (MLIR) infrastructure has emerged as the modern standard for compiler-driven tensor optimization~\cite{mlircgo21}. 
Specifically, the MLIR \texttt{SparseTensor} dialect operationalizes the CSF abstraction, lowering these theoretical format annotations directly into executable iteration strategies.

Within this dialect, the physical memory access pattern for a given axis is dictated entirely by its sparsity annotation. 
If an axis is dense, its structural layout is implicitly defined by its fixed dimension size. 
To traverse it, MLIR generates a standard sequential loop iterating from zero to the dimension's upper bound. 
Because all elements along a dense axis are assumed to exist within the coordinate space, the physical location of any element in the underlying storage can be computed mathematically. 
This implicit structure allows MLIR to perform random-access lookups in constant time.

Conversely, if an axis is compressed, it explicitly stores only the non-zero elements using auxiliary data structures, typically a position (or pointer) array and a coordinate (or index) array. 
Iteration along a compressed axis inherently requires indirect memory access. 
To traverse the non-zeros belonging to a specific parent node in the CSF tree, the MLIR generates a loop that reads a bounding segment from the position array (\eg from $pos[p]$ to $pos[p+1]$) and iterates strictly within those bounds. 
During this traversal, the physical logical coordinate of each non-zero must be explicitly fetched from the coordinate array. 
Because the logical tensor coordinate is decoupled from its physical storage location, direct $O(1)$ random access is impossible. 
Instead, the MLIR must sequentially traverse, coiterate, or binary-search these auxiliary arrays to locate specific elements and perform computations. 

\subsection{Sparse Coiteration} \label{subsec:sparsecoiteration}

\begin{figure*}[t]
	\vspace{-2em}
    \centering
    \begin{subfigure}[b]{0.13\textwidth}
        \centering
        \includegraphics[width=\linewidth]{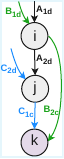}
        \caption{Iteration graph.}
        \label{fig:spgemm_graph}
    \end{subfigure}
    \hfill
    \begin{subfigure}[b]{0.32\textwidth}
        \centering
\begin{lstlisting}[style=acm_mlir_style, language=MLIR, moredelim={[is][\color{gray!95}]{|}{|}}, moredelim={[is][\itshape\color{acmgreen}]{~}{~}}]
|for (int i = 0; i < dim_i; ++i){|
|  for (int j = 0; j < dim_j; ++j){|
    |// Load dim-k pointer|
    |double acc = 0.0;|
    while (pB < ... && pC < ...){
        int kB = B_col[pB];
        int kC = C_row[pC];
        if (kB == kC) {
          acc += B_val[pB] * C_val[pC];
          pB++; pC++;
        } else
        (kB < kC) ? pB++ : pC++;
    }
    A[i][j] = acc;
}}
\end{lstlisting}
        \caption{Two-finger loop in C.}
        \label{fig:spgemm_c}
    \end{subfigure}
    \hfill
    \begin{subfigure}[b]{0.54\textwidth}
        \centering
\begin{lstlisting}[style=acm_mlir_style, language=MLIR, moredelim={[is][\color{gray!95}]{|}{|}}, moredelim={[is][\itshape\color{acmgreen}]{~}{~}}]
|scf.for %i = %c0 to %dim_i step %c1|
|  scf.for %j = %c0 to %dim_j step %c1|
  %res:3 = scf.while(%pB = ..., %pC = ..., %acc = ...){
      scf.condition(~/* %pB < ... && %pC < ... */~) %pB, %pC, %acc
    } do {
    ^bb0(%pB: index , %pC: index , %acc: f64):
      %kB, %kC = // memref.load indices for %pB and %pC 
      %next_acc = scf.if (arith.cmpi eq, %kB, %kC) -> (f64) {
          // Predicated arithmetic: A(i,j) += B(i,k) * C(k,j)
          scf.yield %new_sum : f64
      } else { scf.yield %acc : f64 }
      // Pointer advancing: Increment %pB or %pC based on kB < kC
      scf.yield %next_pB, %next_pC, %next_acc : index, index, f64
    }
\end{lstlisting}
        \caption{MLIR coiteration lowering resulting in two-finger \texttt{scf.while} loop.}
        \label{fig:spgemm_mlir}
    \end{subfigure}
    \caption{
    For SpGEMM ($A_{ij} = \sum_k {\bf B}_{ik} \cdot {\bf C}_{kj}$) kernel, tensor dimensions are subscripted to indicate dense($d$) vs. compressed($c$). 
    }
    \label{fig:spgemm}
    \vspace{-1em}
\end{figure*}

The true complexity of sparse algebra arises during intersection operations, such as contracting $\textbf{\textit{k}}$-index in SpGeMM. 
As illustrated in Figure~\ref{fig:spgemm_graph}, matrix $\textbf{\textit{B}}$ and $\textbf{\textit{C}}$ both access the $\textbf{\textit{k}}$ dimension.
Crucially, $\textbf{\textit{B$_{2c}$}}$ and $\textbf{\textit{C$_{1c}$}}$ edge labels denote that this dimension is explicitly compressed in both input tensors.
Because random access is impossible on compressed axes, the MLIR cannot emit standard, independent \texttt{for} loops.
Instead it must synthesize a dynamic \textit{coiteration} procedure---commonly referred to as a \textit{two-finger} loop.

To traverse this compressed space, MLIR maintains two independent cursors that track the current non-zeros within their respective coordinate arrays. 
This dynamic \textit{two-finger} coiteration is not an MLIR-specific construct, but rather a fundamental programming pattern universally required for sparse intersection, as mirrored in the equivalent handwritten \texttt{C} routine (Figure~\ref{fig:spgemm_c}).
During each iteration of the compiler-generated \texttt{scf.while} loop (Figure~\ref{fig:spgemm_mlir}), the core tensor arithmetic is strictly predicated on a coordinate match, guarded by an \texttt{scf.if} condition. 
Following this predicated computation, the cursors must advance dynamically based on the relative values of their current coordinates. Rather than emitting complex, nested conditional branches to handle this pointer advancement logic, MLIR optimizes the step using data-dependent \texttt{select} operations. This ensures the cursors increment correctly while minimizing control-flow divergence within the critical path.

Despite these control-flow optimizations, the inherent memory dependencies and the conditional nature of the coiteration loop present a massive barrier to high performance. 
Because loop bounds and execution paths are dictated by runtime values that vary dynamically per cycle, classical static compiler transformations---such as loop tiling, array expansion, and, most critically, standard auto-vectorization---are severely restricted or rendered entirely inapplicable~\cite{williams2009roofline, nuzman2006auto}.

\subsection{Microarchitectural Bottlenecks}

The performance of coiteration is fundamentally constrained by a three-fold conflict with modern microarchitecture. 
First, the irregular memory layout of compressed storage shatters the data-level regularity required for effective SIMD vectorization. 
Because matching coordinates must be discovered via runtime scalar comparisons, traditional compilers fail to safely populate wide vector registers, leaving high-throughput ALU resources underutilized~\cite{intel_simd, Kuhn2023Sep, hardware_accel_sparse}.

Second, the data-dependent nature of sparse intersections creates a branch prediction wall. 
In deep Out-of-Order (OoO) engines, the irregular sparsity patterns result in frequent branch mispredictions, triggering pipeline flushes that incur penalties of 15-20 cycles per occurrence~\cite{comp_ar, seznec_branchpred, fog_microarchitecture}. 
These penalties often dwarf the latency of the actual floating-point computation, rendering the arithmetic units secondary to the cost of control-flow recovery. 

Finally, the Instruction-Level Parallelism (ILP) is restricted by the \textit{two-finger} coiteration loop through inescapable data dependency chains. 
In this scalar model, pointer advancement is strictly coupled to intersection results, and arithmetic is predicated on pointer states (Refer Figure~\ref{fig:spgemm_c}).
This sequentiality serializes execution and starves parallel hardware ports, crippling overall throughput~\cite{comp_ar, wall1991limits, ilp_multipro}.

Optimizing these kernels requires a fundamental decoupling of memory retrieval from intersection evaluation. 
To break these scalar dependency chains and reclaim unutilized SIMD capacity, we must move beyond incremental loop transformations toward a hardware-aware, branchless data-flow model. 
The following section introduces \name, an MLIR optimization pass that addresses these bottlenecks by prepending a vectorized, branchless \textit{fast-lane} to the standard coiteration pipeline.

\section{Design} \label{sec:approach}

The primary objective of \name is to safely vectorize the data-dependent \texttt{scf.while} coiteration loops generated during sparse tensor lowering. 
However, this vectorization opportunity is not universally applicable across all sparse tensor contractions. 
Generating a vectorizable two-finger coiteration loop necessitates that several strict structural preconditions are met.

\subsection{Structural Preconditions} \label{sec:structural_prerequisites}

To successfully identify and optimize the dynamic two-finger coiteration bottleneck, our proposed MLIR pass explicitly targets loop nests that satisfy two strict structural preconditions:

\begin{itemize}
    \item \textbf{Compression of the Contracting Index:} The shared index variable must be explicitly annotated as \textit{compressed} across all interacting input tensors. A representative example is the index $k$ in the SpGEMM kernel (Figure \ref{fig:spgemm}), which is compressed in both $\textbf{\textit{B}}[i,k]$ and $\textbf{\textit{C}}[k,j]$. If this shared dimension is uncompressed (dense) in even a single operand, the compiler avoids coiteration entirely. Instead, it generates a standard \texttt{scf.for} loop to iterate exclusively over the sparse non-zeros while performing direct, $O(1)$ random-access lookups into the dense tensor.
    
    \item \textbf{Innermost Loop Structure:} The generated \texttt{scf.while} coiteration must form the innermost structure of the loop nest. Topologically, this dictates that the shared compressed dimension must be the innermost (fastest-changing) index for all interacting tensors, a configuration clearly depicted in Figure \ref{fig:spgemm}. If the \texttt{scf.while} encloses an inner \texttt{scf.for} loop (\eg iterating over a dense innermost dimension), custom vectorization of the coiteration is unnecessary, as the standard compiler pipeline can readily auto-vectorize the innermost \texttt{for} loop.
\end{itemize}

\subsection{The Dual-Path Vectorization Strategy}

\begin{figure}[t]
    \centering
    \includegraphics[width=1\linewidth]{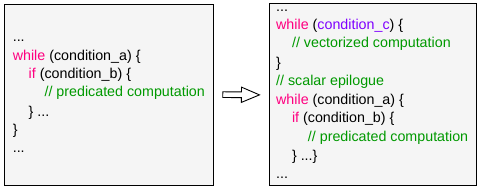}
    \caption{Transforming scalar code into \name dual-path.}
    \label{fig:vectorized}
\end{figure}

To address the microarchitectural bottlenecks inherent in dynamic scalar coiteration, we propose a novel MLIR optimization pass that fundamentally restructures the standard \textit{two-finger} loop pattern.
Specifically, the pass detects instances where \texttt{scf.while} loop forms the innermost structure and relies on divergent \texttt{if-else} control flow to evaluate coordinate intersections (as previously detailed in Section \ref{sec:structural_prerequisites} and Figure \ref{fig:spgemm}).
As illustrated on the left side of Figure~\ref{fig:vectorized}, this baseline scalar loop is governed by a primary bounds check (\texttt{condition\_a}).
For a SpGEMM kernel with cursors $p_B$ and $p_C$ and segment bounds $end_B$ and $end_C$, this global check is defined as ($p_B<end_B \land p_C<end_C$), ensuring that both sparse iterators have remaining elements in their respective segments.
Within the loop, actual computation is predicated on an internal coordinate match (\texttt{condition\_b}).
Figure \ref{fig:spgemm_mlir} provides the complete MLIR lowering for the SpGEMM kernel, explicitly highlighting this critical \textit{two-finger} loop.

Upon identifying this pattern, \name transforms this control flow into a dual-path execution model (Figure~\ref{fig:vectorized}, right).
First, it prepends a vectorized \textit{fast lane}.
Utilizing the MLIR \vectordialect dialect, this new primary loop executes fixed-width SIMD operations.
Based on the target hardware's vector width ($n$), the compiler loads $n$ contiguous non-zero values from each input tensor into vector registers, using vector masks to pad mismatched coordinates with explicit zero values.
Crucially, this fast lane is guarded by a stricter, data-dependent runtime condition (\texttt{condition\_c}).
The augmented vectorization constraint, \texttt{condition\_c}, ensures that a full hardware vector width of elements remains available for SIMD loading---specifically requiring that ($p_B + n \leq end_B \land p_C + n \leq end_C$). 
This check guarantees that the fast lane only executes as long as both tensors possess at least $n$ remaining non-zero elements along the active axis.

The original scalar \scfwhile loop is not entirely eliminated by \name; it is retained to serve as a slow lane scalar epilogue. 
Once the iterators advance near the end of the segment, and the remaining elements fall below the hardware vector width ($n$), \texttt{condition\_c} evaluates to false.
Execution then seamlessly falls through to the unmodified scalar loop, which processes the remaining tail elements using the standard coordinate-matching procedure. 
This guarantees correct algorithmic output without introducing vectorization overhead on small residual segments.

\subsection{Execution Phases}

Mechanically, the execution of the vectorized fast lane is divided into three distinct phases: loading the data, performing the arithmetic computation, and advancing the coordinate pointers. 
To precisely illustrate these phases, the following section continues to use SpGEMM as a running example, utilizing its mathematical formulation and MLIR sparse encoding map (the attribute that explicitly binds logical tensor axes to physical storage levels, such as dense or compressed) to construct and analyze the optimized \texttt{scf.while} innermost loop structure.

Consider the standard SpGEMM kernel computing $\textbf{\textit{A}}[i,j]=\textbf{\textit{B}}[i,k] \cdot \textbf{\textit{C}}[k,j]$. 
To force the compiler to generate the target coiteration, we define specific MLIR sparse encodings for the input tensors:

\begin{itemize}
    \item \textbf{Matrix B} ($\textbf{B}[i,k]$): Encoded in the standard Compressed Sparse Row (CSR) format. The outer dimension $i$ is annotated as \textit{dense}, while the inner contracting dimension $k$ is annotated as \textit{compressed}.
    \item \textbf{Matrix C} ($\textbf{C}[k,j]$): Encoded in the Compressed Sparse Column (CSC) format, mapping $j$ as the outer dense dimension and $k$ as the inner compressed (fastest-changing) dimension. 
    This column-wise access ensures the shared contracting dimension $k$ is physically compressed in both operands, facilitating coiteration.
\end{itemize}

By configuring the compiler to schedule the loop nest in an (${\bf i}, {\bf j}, {\bf k}$) order, the workload decomposes into computing the intersection between the $i$-th row of $\textbf{B}$ and the $j$-th column of $\textbf{C}$. 
Because the shared contracting index $k$ resides at the innermost compressed level for both input tensors, $O(1)$ random access is impossible. 
Consequently, the MLIR sparsifier must emit our exact target code pattern: an innermost \texttt{scf.while} loop that advances two independent pointers to coiterate through the explicitly stored $k$-coordinate arrays. 
As explicitly illustrated in Figure \ref{fig:spgemm}, this dynamically bounded, branch-heavy scalar loop serves as the precise microarchitectural bottleneck that our optimization pass fundamentally restructures.

\paragraph{Phase 1: Contiguous Loading and Mask Generation}
Because the fast lane is strictly guarded by the data-dependent \texttt{condition\_c} (ensuring at least $n$ non-zeros remain in both tensors), the compiler can safely bypass complex gather operations for the initial coordinate retrieval.
Instead, the pass emits standard \texttt{vector.load} operations to fetch $n$ contiguous coordinates from the explicitly stored $k$-index arrays of both $\textbf{B}$ and $\textbf{C}$.

\begin{figure}[t]
\vspace{-0.8em}
\centering
\begin{lstlisting}[style=acm_mlir_style, language=MLIR]
// Original Scalar Data Load
...
%b0_coord = memref.load %B_col_index[%arg1]
%c0_coord = memref.load %C_row_index[%arg2]
%cond = // Calculates if %b0_coord == %c0_coord
%contr = scf.if %cond -> (f64) {
    %b0_val = memref.load %B_values[%arg1]
    %c0_val = memref.load %C_values[%arg2]
    // Tensor contraction
}
...
\end{lstlisting}
\vspace{-1em}
\[
\Downarrow \text{Lowering to Vector Data Loads}
\]
\vspace{-1em}
\begin{lstlisting}[style=acm_mlir_style, language=MLIR]
// After: 4-wide Vector Loads
...
%b_4_coords = vector.load %B_col_index[%arg1]
%c_4_coords = vector.load %C_row_index[%arg2]
%b_4_values = vector.load %B_values[%arg1]
%c_4_values = vector.load %C_values[%arg2]
...
\end{lstlisting}

\caption{Scalar memory loads from sparse tensors lowered to fixed-width \texttt{vector.load} operations.}
\label{fig:vector_load_lowering}
\vspace{-1em}
\end{figure}

Figure \ref{fig:vector_load_lowering} illustrates this load transformation. 
In the baseline scalar execution, the loop fetches individual coordinates from $\textbf{B}$ and $\textbf{C}$, and the actual numerical values are loaded strictly after a conditional \texttt{if-else} branch confirms a coordinate match. 
In contrast, our vectorized fast lane unconditionally fetches $n$ coordinates and $n$ corresponding numerical values simultaneously from both matrices. 
By decoupling the memory loads from the intersection, this transformation entirely eliminates control-flow divergence within the program's hot-path.

\begin{figure}[t]
\centering
\small
\setlength{\tabcolsep}{5pt}
\begin{tabular}{c|cccc}
 & $k_j^{(0)}$ & $k_j^{(1)}$ & $k_j^{(2)}$ & $k_j^{(3)}$ \\
\hline
$k_i^{(0)}$ & $k_i^{(0)} = k_j^{(0)}$ & $k_i^{(0)} = k_j^{(1)}$ & $k_i^{(0)} = k_j^{(2)}$ & $k_i^{(0)} = k_j^{(3)}$ \\
$k_i^{(1)}$ & $k_i^{(1)} = k_j^{(0)}$ & $k_i^{(1)} = k_j^{(1)}$ & $k_i^{(1)} = k_j^{(2)}$ & $k_i^{(1)} = k_j^{(3)}$ \\
$k_i^{(2)}$ & $k_i^{(2)} = k_j^{(0)}$ & $k_i^{(2)} = k_j^{(1)}$ & $k_i^{(2)} = k_j^{(2)}$ & $k_i^{(2)} = k_j^{(3)}$ \\
$k_i^{(3)}$ & $k_i^{(3)} = k_j^{(0)}$ & $k_i^{(3)} = k_j^{(1)}$ & $k_i^{(3)} = k_j^{(2)}$ & $k_i^{(3)} = k_j^{(3)}$ \\
\end{tabular}
\caption{
Naive cartesian coiteration for SpGEMM: All-pairs of the nonzero indices in row $B_i$ and column $C_j$ are compared to identify matching $k$ indices contributing to $A_{ij}$.
}
\label{fig:spgemm_cartesian}
\vspace{-1.5em}
\end{figure}

Once these coordinate vectors populate the SIMD registers, the pass executes an all-pairs cross-comparison strategy, yielding an $n\times n$ boolean matrix representing all potential intersections (illustrated in Figure \ref{fig:spgemm_cartesian}). 
Because the coordinate arrays in CSF formats are strictly increasing, this comparison guarantees at most one \texttt{true} value per row.
This inherent property ensures that there are no duplicate coordinates within either loaded vector; therefore, a specific $k$-index from one tensor can match, at most, a single identical $k$-index in the other.

In the context of our running example SpGEMM (computing the intersection of the $\textbf{i}$-th row of $\textbf{B}$ and the $\textbf{j}$-th column of $\textbf{C}$), this property ensures that for any specific $\textbf{k}$-coordinate in $\textbf{B}$, there can be at most one matching $\textbf{k}$-coordinate in $\textbf{C}$.
If a row in the cross-comparison mask yields entirely \texttt{false} values, it indicates a strict coordinate mismatch, meaning the specific $\textbf{k}$-index exists in one operand but not the other, and therefore will not contribute to the final tensor contraction.

\begin{figure*}[t]
    \centering
    \includegraphics[width=1\linewidth]{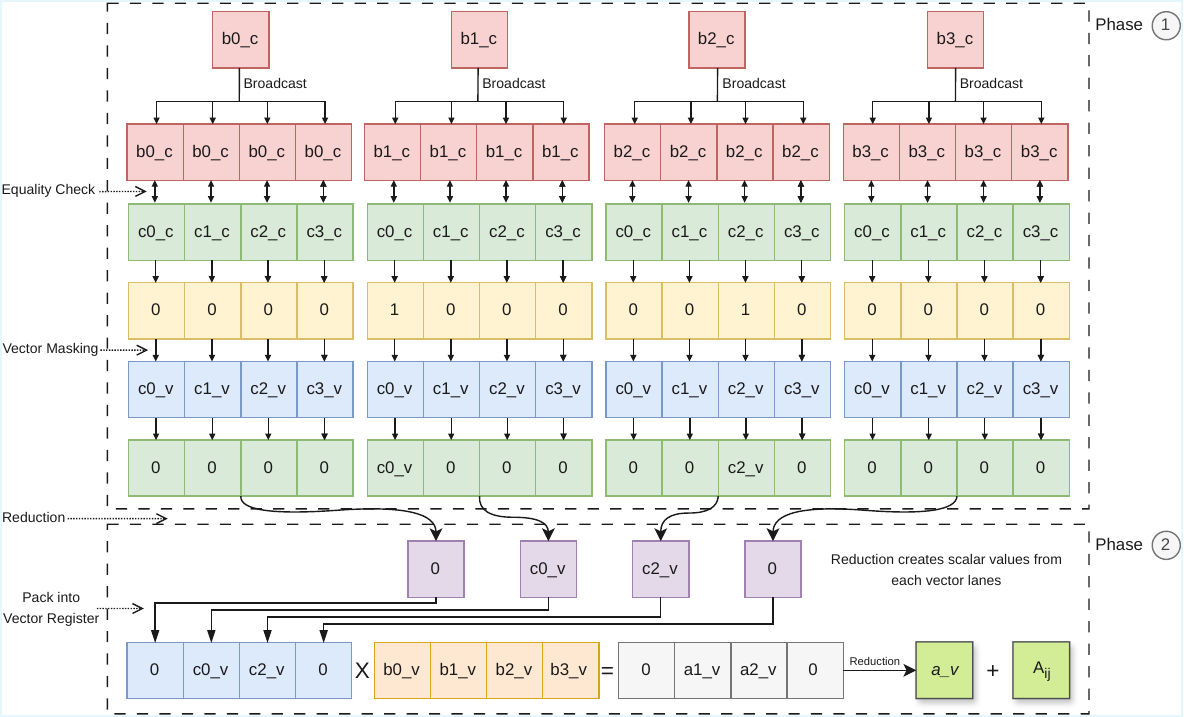}
    \caption{
    Dataflow architecture of a branchless $4\times 4$ SIMD cross-compare algorithm for SpGEMM. 
    The execution model bypasses scalar control-flow by broadcasting matrix $B$ coordinates to perform parallel equality checks against matrix $C$ coordinates. 
    The resulting boolean masks filter matrix $C$ values, which are then horizontally reduced to extract the matched non-zeros (or structural zeros) for the subsequent accumulate stage.
    }
    \label{fig:computation_circuit}
\end{figure*}

As illustrated in the detailed dataflow of Figure \ref{fig:computation_circuit}, the pass implements this all-pairs comparison natively within the SIMD pipeline. 
The $n$ coordinate values from \textbf{B} are individually broadcast into full-width vector registers and checked for equality against the contiguous coordinate vector from \textbf{C}. 
The resulting boolean vectors serve as masks to filter the loaded values of $\textbf{C}$. 
Subsequently, a horizontal reduction extracts the single matched value (or a structural zero) from each masked vector, producing the necessary operands for the core tensor computation.

\paragraph{Phase 2: Predicated Tensor Computation}
Following the cross-comparison and the subsequent extraction of matched values, the pass proceeds to the core arithmetic computation. 
As illustrated at the transition into Phase 2 of Figure \ref{fig:computation_circuit}, the horizontal reduction of the masked vectors yields discrete scalar values. 
To maintain a fully vectorized dataflow pipeline, our initial approach is to repack these discrete values---comprising both matched non-zeros (\eg \texttt{c0\_v}) and explicit zero values---back into a cohesive SIMD register.  
The operation \texttt{vector.from\_elements} helps to achieve repackaging within the MLIR \vectordialect dialect (Figure \ref{fig:splyce_tensor_vec}), effectively constructing a dense vector operand that represents the filtered values of \textbf{C}.

With both operand vectors fully populated, the core arithmetic is executed natively via fixed-width SIMD instructions. 
Because the loop's output accumulator (\ie \texttt{\%Aij}) is a scalar value, the pass cannot utilize a standard fused multiply-add (\texttt{vector.fma}). 
Instead, the semi-ring multiply-accumulate (MAC) decomposes into a two-step vectorized operation. 
First, the pass emits an unconditional element-wise vector multiplication (\texttt{arith.mulf}) between the \textbf{B} and \textbf{C} registers. 
Subsequently, the resulting product vector undergoes a horizontal sum reduction (\texttt{vector.reduction <add>}) to collapse the array into a single scalar value (\texttt{\%a\_v}), which is then added to the loop's running accumulator.

\begin{figure*}[t]
\centering

\begin{subfigure}[b]{0.43\textwidth}
\centering
\begin{lstlisting}[style=acm_mlir_style, language=MLIR]
%contr = scf.if %cond -> (f64) {
    %b0_val = memref.load %B_v[%arg1]
    %c0_val = memref.load %C_v[%arg2]
    %mul = arith.mulf %b0_val, %c0_val
    %new_acc = arith.addf %Aij, %mul
}
\end{lstlisting}
\caption{Baseline scalar execution generated from MLIR lowering.}
\label{fig:mlir_tensor_scalar}
\end{subfigure}
\hfill
\begin{subfigure}[b]{0.54\textwidth}
\centering
\begin{lstlisting}[style=acm_mlir_style, language=MLIR]
// Pack C values into a vector
%c_vals_vec  = vector.from_elements %c0_v, %c1_v, %c2_v, %c3_v        
%contrib_vec = arith.mulf %b_vals_vec, %c_vals_vec 
// Horizontal reduction
%a_v = vector.reduction <add>, %contrib_vec
%new_acc = arith.addf %Aij, %a_v
\end{lstlisting}
\caption{\name-vectorized branchless SIMD computation and reduction.}
\label{fig:splyce_tensor_vec}
\end{subfigure}

\caption{Transform of the core computation in Phase 2.}

\label{fig:vector_contraction}
\vspace{-1em}
\end{figure*}

Crucially, to sustain peak instruction throughput, this vectorized computation is executed entirely without branching. 
Unlike the baseline MLIR lowering (Figure~\ref{fig:mlir_tensor_scalar}), which strictly guards scalar computation with an \texttt{scf.if} branch to process only true non-zeros, our transformed fast lane unconditionally executes the SIMD arithmetic across the entire vector width. 
While this approach inherently introduces redundant arithmetic---such as computing $b0\_v \times 0 = 0$ on unmatched lanes, as depicted in the Phase 2 multiplication block---inserting conditional extraction logic to bypass these zero values would shatter the SIMD pipeline and induce frequent control-flow divergence. 
On modern superscalar architectures, the latency penalty for branch mispredictions and the resulting instruction pipeline flushes far outweighs the localized cost of executing dummy floating-point operations within the vector hardware.

\paragraph{Phase 3: Dynamic Pointer Advancement}
After the predicated computation completes, the independent iteration cursors for Matrices \textbf{B} and \textbf{C} must be updated to prepare for the next iteration of the \texttt{scf.while} loop. 
Because the sparse coordinate distributions are irregular, the cursors cannot uniformly advance by the hardware vector width $n$. 
If one tensor's loaded coordinates vastly outpace the other's, aggressively advancing both pointers by $n$ would skip valid intersections.

\begin{figure}
    \centering
    \includegraphics[width=0.85\linewidth]{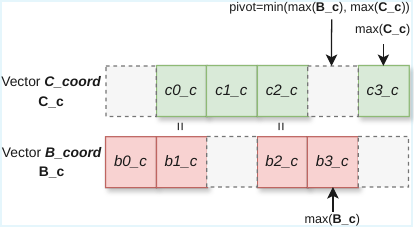}
    \caption{
    Branchless pointer advancement strategy.
    }
    \label{fig:pointer_advancement}
\end{figure}

To theoretically maintain a purely vectorized pipeline across the entire coiteration cycle, our initial approach formulates pointer advancement as a branchless sequence of vector operations. 
The underlying formulation relies on establishing a safe execution boundary, or \textit{pivot}, which represents the largest coordinate up to which both tensors have been fully inspected.
As illustrated in Figure \ref{fig:pointer_advancement}, this pivot is dynamically calculated as the minimum of the maximum coordinates currently loaded in the respective vector chunks (\ie \textit{pivot} $= min(max(B\_coord), max(C\_coord))$). 

Rather than relying on conditional scalar loops to evaluate the next step, the compiler calculates the exact cursor advancement natively within the SIMD unit, detailed by the MLIR lowering in Figure \ref{fig:vector_pointer_advancement}.
The scalar pivot is broadcast into a vector register and evaluated against the original loaded coordinate vectors via a parallel less-than-or-equal-to comparison. 
This yields a boolean mask in which true lanes indicate elements that have been safely consumed.

To quantify the final advancement step, these boolean results are zero-extended into integers (converting true to 1 and false to 0) and summed using a horizontal vector reduction. 
The resulting scalar unconditionally increments the tensor's base pointer for the next loop iteration. 
By formulating this advancement entirely through vector comparisons and reductions, the methodology achieves dynamic, data-dependent cursor updates without introducing a single control-flow branch.

\subsection{Maximizing Parallel Throughput}

While the purely vectorized, branchless architecture proposed above theoretically maximizes SIMD lane utilization, practical hardware constraints dictate that aggressive vectorization is not universally optimal. 
Specifically, the latency overheads associated with dynamically extracting matching values from vectors, repacking discrete values into new vector registers, and executing cross-lane shuffles can frequently eclipse the throughput gains of parallel execution.

To mitigate these potential microarchitectural bottlenecks, the $n$ non-zeros processed in a single fast-lane cycle can alternatively be handled via highly unrolled scalar instructions. 
For all three phases discussed, it is theoretically possible to bypass the SIMD pipeline entirely and emit discrete scalar operations:

\begin{itemize}
    \item \textbf{Phase 1}: Instead of vectorized masking and reduction to filter non-zeros, the $n\times n$ all-pairs cross-comparison can be executed directly via independent scalar equality checks and loading the matched values straight into scalar registers.
    \item \textbf{Phase 2}: Instead of repacking the extracted matching values back into SIMD registers, the arithmetic can be computed sequentially using unrolled scalar MAC instructions.
    \item \textbf{Phase 3}: The branchless pointer calculation, as shown in Figure \ref{fig:vector_pointer_advancement}, can bypass the vector unit entirely. As demonstrated in Figure \ref{fig:scalar_pointer_advancement}, the pointer advancement can be executed using unrolled scalar comparisons and a tree-reduction.
\end{itemize}

\begin{figure*}[t]
  \centering
  \begin{minipage}[t]{0.48\textwidth}
    \vspace{0pt} 
    \begin{lstlisting}[style=acm_mlir_style, language=MLIR]
%pivot_v    = vector.broadcast %pivot
%b_le_vec   = arith.cmpi ule, %b_c, %pivot_v
%b_le_int   = arith.extui %b_le_vec
%b_step_val = vector.reduction <add>, %b_le_int
%b_step     = arith.index_cast %b_step_val
%new_ptr_B  = arith.addi %ptr_B, %b_step
    \end{lstlisting}
    \caption{Branchless vectorized pointer advancement.}
    \label{fig:vector_pointer_advancement}
  \end{minipage}
  \hfill
  \begin{minipage}[t]{0.48\textwidth}
    \vspace{0pt}
    \begin{lstlisting}[style=acm_mlir_style, language=MLIR]
%b0_le = arith.cmpi ule, %b0, %pivot : index ...
%b0_adv = arith.select %b0_le, %c1, %c0 : index ...
%b_s01 = arith.addi %b0_adv, %b1_adv : index
%b_s23 = arith.addi %b2_adv, %b3_adv : index
%b_step = arith.addi %b_s01, %b_s23 : index
%new_ptr_B = arith.addi %ptr_B, %b_step : index
    \end{lstlisting}
    \caption{Unrolled scalar variant of pointer advancement.}
    \label{fig:scalar_pointer_advancement}
  \end{minipage}
\vspace{-1.5em}
\end{figure*}


Because of these competing microarchitectural trade-offs, the optimal configuration for the fast lane is a finely-tuned hybrid execution model.
Rather than relying on theoretical assumptions or runtime auto-tuning, the architectural design of our final compiler pass is strictly informed by hardware realities---specifically, how the CPU's Out-of-Order execution engine and ILP interact with vector packing latencies.
We conducted an ablation study to determine exactly which phases of our pass must be vectorized.

\section{Evaluation} \label{sec:evaluation}

In this section, we evaluate the performance, scalability, and microarchitectural behavior of our proposed dual-path MLIR optimization pass. Our experimental analysis is designed to answer four primary research questions:

\begin{itemize}
\item \textbf{RQ1 (Phase Ablation):} Among data loading, tensor computation, and pointer advancement, does vectorizing all three execution phases yield the optimal microarchitectural performance?
\item \textbf{RQ2 (Vectorization Overhead):} Does the performance benefit of vectorizing sparse coiteration via SIMD instructions outweigh the computational overhead of processing explicit zero values?
\item \textbf{RQ3 (Sparsity Scaling):} Does the proposed dual-path architecture maintain its performance advantages across varying tensor dimensions and data densities?
\item \textbf{RQ4 (Parallel Scalability):} Are the ILP gains achieved by \name orthogonal to macro-scale thread-level parallelism (TLP) when scaled across parallel multicore environments?
\end{itemize}

All experiments were conducted on a dual-socket Intel Xeon Gold 6430 (Sapphire Rapids) server. 
The system features a 2-node NUMA architecture with a total of 64 physical cores (32 per socket). 
To isolate microarchitectural behavior and ensure deterministic, cycle-accurate profiling, Simultaneous Multithreading (SMT) was disabled, and all benchmark threads were strictly pinned to a single NUMA node. 
The processors feature 120\,MB of shared L3 cache and 2\,MB of dedicated L2 cache per core, alongside native AVX-512 instruction support. 
The host operating system is Ubuntu 24.04.4 LTS (Linux kernel 6.x). 
The \name optimization pass is
implemented and evaluated against the LLVM/MLIR \texttt{23.0.0-git} mainline. 

All benchmark kernels were compiled with strict performance flags: \texttt{-O3 -march=native}.
Although the \texttt{-Ofast} compiler flag enables floating-point reassociation\textemdash allowing the compiler to break the loop-carried dependency chain in multiply-accumulate operations that would otherwise negate vectorization gains\textemdash the default MLIR sparsifier pipeline does not enable \textit{fast-math} optimizations.
Consequently, \name inherits these conservative floating-point semantics by default throughout our evaluation.

\subsection{Phase Ablation}
\begin{figure}[h]
    \centering
    \includegraphics[width=\linewidth]{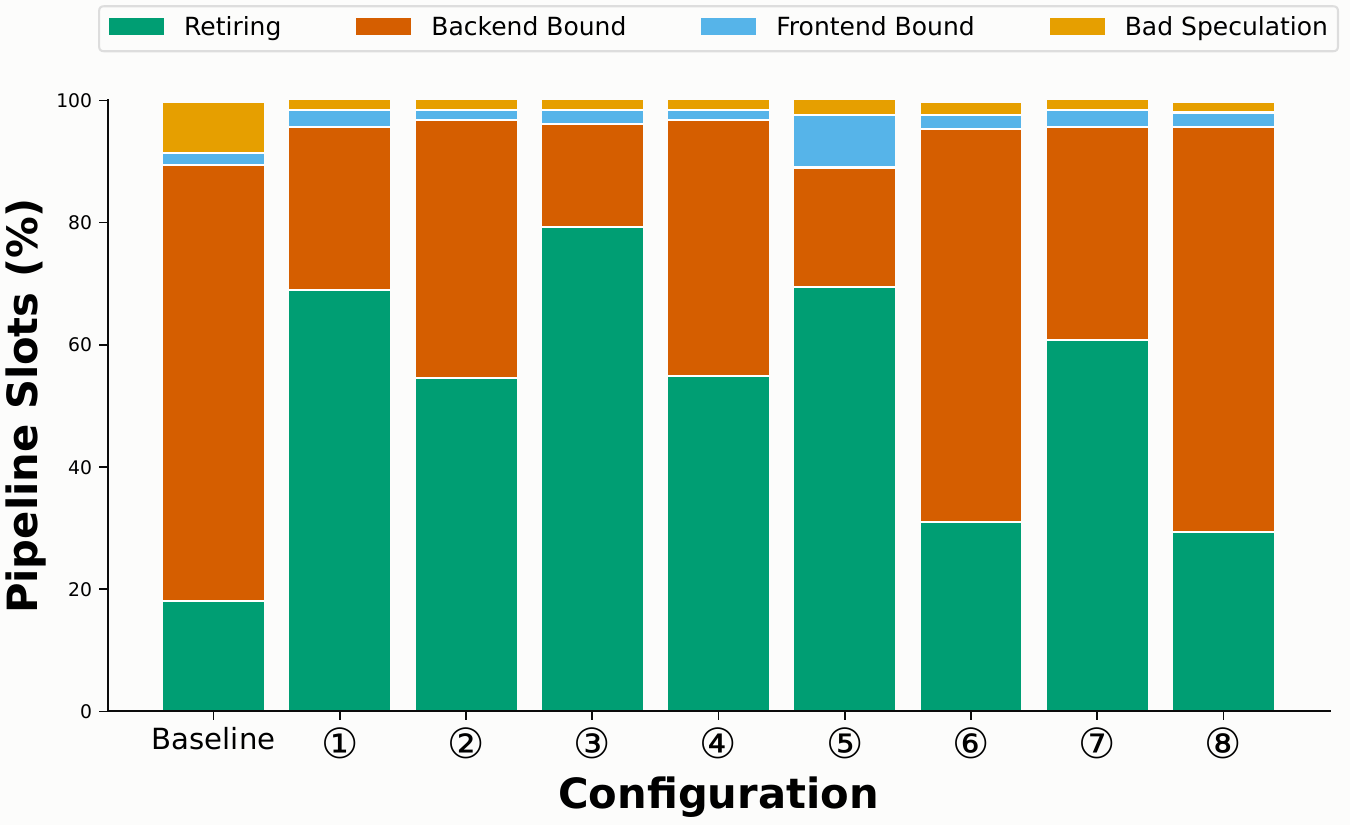}
    \caption{Top-Down Microarchitecture Analysis (TMA) pipeline slot utilization across different configurations.}
    \label{fig:tma_results}
\end{figure}

To address \textbf{RQ1} and determine the optimal microarchitectural configuration, we conducted a rigorous ablation study on the SpGEMM fast lane. 
By selectively enabling SIMD vectorization across the three execution phases, we isolated the performance impact of each component. 
The results, detailed in Table~\ref{tab:spgemm_ablation} and Figure~\ref{fig:tma_results}, reveal a critical insight: full-pipeline vectorization does not yield the optimal microarchitectural performance.

\begin{table}[h]
    \centering
    \caption{Ablation study of SpGEMM coiteration phases (\checkmark: SIMD, $\times$: scalar). Baseline (58.57s) represents the unmodified MLIR sparsifier.}
    \label{tab:spgemm_ablation}
    
    \small
    \setlength{\tabcolsep}{5.2pt} 
    \renewcommand{\arraystretch}{1.0} 
    \begin{tabular}{@{} c ccc SSSS r @{}}
    \toprule
    & \multicolumn{3}{c}{\textbf{Phases}} & \multicolumn{1}{c}{\textbf{Insn.}} & & \multicolumn{1}{c}{\textbf{Br. Miss}} & \multicolumn{1}{c}{\textbf{Time}} & \\
    \cmidrule(lr){2-4}
    \textbf{\#} & \textbf{1} & \textbf{2} & \textbf{3} & \multicolumn{1}{c}{\scriptsize($10^{11}$)} & \multicolumn{1}{c}{\textbf{IPC}} & \multicolumn{1}{c}{\scriptsize($10^{7}$)} & \multicolumn{1}{c}{\scriptsize(s)} & \multicolumn{1}{c}{\textbf{Speedup}} \\
    \midrule
    -- & \multicolumn{3}{c}{Original MLIR} & 1.59739608625 & 1.306 & 34.1004592 & 58.565833 & $1.00\times$ \\
    \midrule
    \confnum{1} & $\times$ & $\times$ & $\times$ & 2.71550812702 & 3.946 & 5.3232963 & 33.014631 & $1.77\times$ \\
    \confnum{2} & $\times$ & $\times$ & \checkmark & 2.96835758394 & 3.137 & 5.3372894 & 45.358299 & $1.29\times$ \\
    \confnum{3} & $\times$ & \checkmark & $\times$ & 2.92628971175 & 4.542 & 5.3143079 & 30.911874 & $1.89\times$ \\
    \confnum{4} & $\times$ & \checkmark & \checkmark & 3.05256540732 & 3.238 & 5.3347971 & 45.193768 & $1.30\times$ \\
    \rowcolor{gray!15}
    \confnum{5} & \checkmark & $\times$ & $\times$ & 1.59821159013 & 3.639 & 5.3157819 & \textbf{21.10} & \textbf{2.78}$\times$ \\
    \confnum{6} & \checkmark & $\times$ & \checkmark & 1.45147746487 & 1.729 & 5.3172118 & 40.252762 & $1.46\times$ \\
    \confnum{7} & \checkmark & \checkmark & $\times$ & 1.68241944790 & 3.314 & 5.3124763 & 24.371618 & $2.40\times$ \\
    \confnum{8} & \checkmark & \checkmark & \checkmark & 1.53568531716 & 1.716 & 5.3193017 & 42.908178 & $1.36\times$ \\
    \bottomrule
    \end{tabular}
\end{table}

 \label{sec:phase_ablation}

Before analyzing the SIMD scaling, the data demonstrates the inherent structural advantage of the dual-path model itself. 
Compared to the unmodified MLIR baseline, every dual-path configuration achieves an approximate $6.41\times$ reduction in branch mispredictions. 
This is the direct consequence of eliminating the divergent \texttt{if-else} coordinate matching logic from the \texttt{scf.while} hot path. 
By migrating from unpredictable data-dependent branches to bulk cross-comparisons, the architecture drastically reduces pipeline flushes. 
Figure~\ref{fig:tma_results}, a Level 1 Top-Down Microarchitecture Analysis (TMA)~\cite{tmayasin2014} chart detailing all phase configurations alongside the MLIR baseline, visually corroborates this by demonstrating a strict reduction in \textit{Bad Speculation} pipeline slots.

However, while this structural transformation mitigates control-flow penalties, attempting to vectorize the pointer advancement (Phase 3) introduces a fatal execution bottleneck. 
As explicitly shown in Table~\ref{tab:spgemm_ablation}, the lowest overall speedups (\confnum{2}, \confnum{4}, \confnum{6}, and \confnum{8}) occur precisely when Phase 3 is vectorized. 
Consequently, as illustrated in Figure~\ref{fig:tma_results}, these specific configurations suffer a massive spike in \textit{Backend Bound} stalls.
This indicates that the CPU is stalling not due to memory latency or frontend starvation, but because its backend execution units are severely overwhelmed. 
Advancing dynamically bounded sparse pointers in SIMD not only requires expensive cross-lane shuffles, but it also triggers a severe microarchitectural hardware clash. 
The CPU's Address Generation Unit (AGU) requires memory addresses to reside in scalar General Purpose Registers (GPRs). 
When pointer arithmetic is forced into vector registers---either by design or by the LLVM SLP auto-vectorizer under \texttt{-O3}---the hardware must execute high-latency domain-crossing instructions (\eg \texttt{vpextrq}) or serialized Gather operations to feed the AGU for the subsequent iteration. 
To bypass this compiler cost-model myopia and preserve optimal AGU throughput, Phase 3 must remain strictly scalar, necessitating explicit \texttt{-fno-slp-vectorize} and \texttt{-fno-vectorize} compiler flags.

Beyond the penalties of Phase 3, the ablation data highlights the absolute necessity of vectorizing the memory loads and coordinate cross-comparisons (Phase 1). 
Configurations where Phase 1 remains scalar exhibit a massive inflation in total executed instructions. 
Notably, \confnum{3} achieves the highest Instructions Per Cycle (IPC) across all configurations, yet yields a suboptimal execution time. 
This empirical data demonstrates a classic SIMD IPC paradox: the scalar coiteration loop's high IPC consists of billions of lightweight scalar coordinate comparisons (an $n \times n$ matching complexity) that keep the pipeline highly active but yield little semantic progress towards the high-level tensor operation.
Vectorizing Phase 1 collapses this complexity, mathematically deflating the total instruction count via bulk SIMD comparisons and proving mandatory for achieving significant speedup.

Having established the necessity of a vectorized Phase 1 and a scalar Phase 3, the final architectural decision rests on Phase 2 (Predicated Tensor Computation). 
Table~\ref{tab:spgemm_ablation} confirms that \confnum{5}, which executes Phase 2 using standard scalar instructions, achieves the absolute peak speedup ($2.78\times$). 
This superiority stems from a microarchitectural trade-off between SIMD packing latencies and ILP. 
To execute Phase 2 in SIMD, the hardware must first repack the dynamically matched sparse values into contiguous vector registers---a process strictly gated by the Phase 1 vector mask. 
This dynamic permutation overhead severely degrades throughput.
Conversely, emitting independent scalar MAC instructions allows the modern OoO execution engine to naturally exploit ILP, completely bypassing the SIMD repack latencies. 
Thus, avoiding vectorization overhead in Phase 2 in favor of hardware-driven ILP ultimately yields the optimal performance.
Because this optimal configuration is highly dependent on underlying hardware capabilities, \name exposes an optional compile-time flag to independently toggle the vectorization of each phase.

\begin{table}[t]
    \centering
    \caption{Ablation study of SpGEMM coiteration phases with associative math enabled in Splyce (\checkmark: SIMD, $\times$: scalar). Baseline (58.57s) represents the unmodified MLIR sparsifier.}
    \label{tab:spgemm_ablation_fastmath}
    
    \small
    \setlength{\tabcolsep}{5.2pt} 
    \renewcommand{\arraystretch}{1.0} 
    \begin{tabular}{@{} c ccc SSSS r @{}}
    \toprule
    & \multicolumn{3}{c}{\textbf{Phases}} & \multicolumn{1}{c}{\textbf{Insn.}} & & \multicolumn{1}{c}{\textbf{Br. Miss}} & \multicolumn{1}{c}{\textbf{Time}} & \\
    \cmidrule(lr){2-4}
    \textbf{\#} & \textbf{1} & \textbf{2} & \textbf{3} & \multicolumn{1}{c}{\scriptsize($10^{11}$)} & \multicolumn{1}{c}{\textbf{IPC}} & \multicolumn{1}{c}{\scriptsize($10^{7}$)} & \multicolumn{1}{c}{\scriptsize(s)} & \multicolumn{1}{c}{\textbf{Speedup}} \\
    \midrule
    -- & \multicolumn{3}{c}{Original MLIR} & 1.59739608625 & 1.306 & 34.1004592 & 58.565833 & $1.00\times$ \\
    \midrule
    \confnum{1}-f & $\times$ & $\times$ & $\times$ & 2.73679017020 & 4.537 & 5.3164538 & 32.96 & $1.78\times$ \\
    \confnum{2}-f & $\times$ & $\times$ & \checkmark & 2.90516978132 & 3.195 & 5.3188636 & 43.593232 & $1.34\times$ \\
    \confnum{3}-f & $\times$ & \checkmark & $\times$ & 2.88415388571 & 4.447 & 5.3122681 & 31.113477 & $1.88\times$ \\
    \confnum{4}-f & $\times$ & \checkmark & \checkmark & 3.01042958184 & 3.170 & 5.3415574 & 45.524428 & $1.29\times$ \\
    \rowcolor{gray!15}
    \confnum{5}-f & \checkmark & $\times$ & $\times$ & 1.36660814968 & 3.206 & 5.3053809 & \textbf{20.48} & \textbf{2.86}$\times$ \\
    \confnum{6}-f & \checkmark & $\times$ & \checkmark & 1.21987400304 & 1.560 & 5.3200975 & 37.501559 & $1.56\times$ \\
    \confnum{7}-f & \checkmark & \checkmark & $\times$ & 1.47186793300 & 3.429 & 5.3182320 & 20.619093 & $2.84\times$ \\
    \confnum{8}-f & \checkmark & \checkmark & \checkmark & 1.32513378836 & 1.676 & 5.3282829 & 37.906000 & $1.54\times$ \\
    \bottomrule
    \end{tabular}
\end{table}

To preserve strict IEEE 754 semantics, the default MLIR sparsifier pipeline prohibits floating-point reassociation, which inherently restricts the potential for tree-height reduction. 
To explore this optimization space, \name introduces an optional flag that explicitly enables associative math transformations. 
As detailed in Table~\ref{tab:spgemm_ablation_fastmath}, activating this flag accelerates execution across all eight phase configurations relative to the original MLIR program. 
However, to uphold standardized floating-point guarantees and align with the default MLIR sparsifier, we isolate this ablation and omit the associative flag in all subsequent experiments.

Beyond the phase-level vectorization strategy, the efficiency of the dual-path model depends heavily on the hardware vector length ($n$), which dictates the number of elements processed per loop iteration. 
For 64-bit double-precision floating-point workloads, the native AVX-512 instruction set allows for a maximum vector width of $n=8$. 
However, as detailed in Figure~\ref{fig:spgemm_vectorsize}, maximizing the vector width to $n = 8$ empirically yields suboptimal performance across various sparsity densities.

\input{figure/spgemm_vectorsize}

This degradation reveals a fundamental microarchitectural bottleneck due to AVX-512 predicate register pressure. 
While the architecture natively supports 8 elements per vector register, it provisions only 7 usable mask registers (\texttt{k1}--\texttt{k7}) for conditional execution, as \texttt{k0} is hardwired to disable masking~\cite{intel_avx512}. 
As the vectorization factor scales to $n=8$, the intermediate boolean operations required to compute the Phase 1 coordinate cross-comparison masks exceed this 7-register limit. 
Consequently, the compiler is forced to spill active predicate masks to standard GPRs and execute costly \texttt{kmov} instructions to shuttle masks in and out of the execution units. 
Coupled with the inherently lower SIMD lane utilization that wider vectors suffer when processing sparse data, this register spilling severely degrades overall throughput.

Recognizing that the optimal vector width is dictated by a delicate balance between hardware register capacity, ISA constraints, and dataset sparsity, we designed our MLIR optimization pass to expose the vectorization factor $n$ as a highly configurable compilation parameter rather than a rigid architectural constant. 
Based on our empirical profiling, setting $n=4$ achieves the optimal microarchitectural trade-off for our target platform; consequently, this configuration is adopted as the standard vector width for all remaining experiments in this evaluation.

\subsection{Vectorization Overhead} \label{sec:vec_overhead}

\begin{table*}[t]
\vspace{-1em}
\centering
\caption{Performance of sparse kernels on synthetic datasets (5\% uniform sparsity). 
Boldface variables in the equations denote sparse tensors, whereas non-bold variables represent dense tensors. 
}
\label{tab:basic_kernels}
\begin{tabular}{lccccc}
\toprule
\textbf{Kernel} & \textbf{Equation} & \textbf{Input Dimension} & \textbf{Baseline (s)} & \textbf{Optimized (s)} & \textbf{Speedup} \\
\midrule
SpGEMM & $A_{ij} = \sum_{k}\textbf{B}_{ik} \cdot \textbf{C}_{kj}$ & $\textbf{B}(5k\!\times\!5k),\, \textbf{C}(5k\!\times\!5k)$ & 58.71 & 20.83 & 2.82$\times$ \\
SpMSpV & $y_{i} = \sum_{j}\textbf{B}_{ij} \cdot \textbf{x}_{j}$ & $\textbf{B}(100\!\times\!100M),\, \textbf{x}(100M)$ & 4.81 & 1.68 & 2.86$\times$ \\
SpMTTKRP & $A_{ik} = \sum_{k,l}\textbf{B}_{ikl} \cdot \textbf{C}_{lj} \cdot \textbf{D}_{kj}$ & $\textbf{B}(1k\!\times\!1k\!\times\!1k),\, \textbf{C}(1k\!\times\!1k),\, \textbf{D}(1k\!\times\!1k)$ & 24.33 & 10.65 & 2.28$\times$ \\
SpMMH & $A_{ij} = \sum_{k}\textbf{B}_{ik} \cdot C_{kj} \cdot \textbf{D}_{ij}$ & $\textbf{B}(5k\!\times\!5k),\, C(5k\!\times\!5k),\, \textbf{D}(5k\!\times\!5k)$ & 63.82 & 32.63 & 1.96$\times$ \\
SpTTSpM & $A_{ijr} = \sum_{k}\textbf{B}_{ijk} \cdot \textbf{C}_{kr}$ & $\textbf{B}(500\!\times\!500\!\times\!500),\, \textbf{C}(500\!\times\!500)$ & 30.53 & 14.39 & 2.12$\times$ \\
\bottomrule
\end{tabular}
\end{table*}

To address \textbf{RQ2} and quantify the computational trade-offs of our approach, we evaluated five sparse tensor contraction kernels: SpGEMM, SpMSpV, SpMTTKRP, SpMMH, and SpTTSpM. 
This initial evaluation used synthetic datasets with a uniform 5\% non-zero values along the innermost compressed dimension. 
As detailed in Table~\ref{tab:basic_kernels}, the proposed dual-path architecture yields end-to-end speedups ranging from $1.96\times$ to $2.86\times$, achieving an overall geometric mean speedup of 2.38$\times$ against the standard MLIR baseline. 
This data explicitly proves our hypothesis: the immense throughput benefits of bulk SIMD coordinate matching and instruction deflation comprehensively outweigh the computational overhead of processing explicit zero-padding within the vector lanes.
Furthermore, to validate the practical applicability of this optimization beyond uniformly distributed synthetic data, we extended our evaluation to highly irregular, real-world sparse tensors sourced from the SuiteSparse Matrix Collection~\cite{suitesparse1, suitesparse2}. 
Across these production-grade datasets, the dual-path fast lane demonstrates improvements across the kernels.

\subsection{Sparsity Scaling} \label{sec:sparsity_scaling}

Addressing \textbf{RQ3} is critical for establishing the operational boundaries of the optimization---specifically, determining the sparsity thresholds that dictate when this optimization pass yields profitable returns. 
To map this behavior, we conducted a sensitivity analysis on the SpGEMM kernel, sweeping the sparsity of the input matrices (uniform random distribution; identical for both operands). 
As illustrated in Figure~\ref{fig:spgemm_runtime_sparsity}, the performance delta between the baseline and the optimized fast lane scales positively with the sparsity of the matrix. 
Crucially, even at an extremely low 0.01\% sparsity regime, the optimized kernel maintains strict execution parity with the unmodified baseline.

\input{figure/spgemm_sparsity}

This trend extends to real-world datasets (Table~\ref{tab:real_tensor_results}).
While many production tensors often possess much lower global sparsity than our synthetic benchmarks, the same convergence toward 1.0$\times$ speedup is observed in the limit, such as with the \texttt{internet} dataset (1.33e-3\% sparsity).
Crucially, these results highlight the performance safety of our dual-path strategy. 
The vectorized fast-lane targets throughput gains within dense clusters, while the scalar epilogue ensures that execution remains consistent with the baseline in the sparse limit. 
In this context, global sparsity is not the sole determining factor, as globally sparse datasets can still contain highly dense local regions.
The SpMSpV kernel results in Table~\ref{tab:real_tensor_results} exemplify this: the highly sparse \texttt{stokes} matrix achieves a significant speedup, whereas \texttt{hugetrace-00020}, which falls in a similar range, shows almost no speedup.
However, the \textit{SIMD(\%)} metric reveals that the portion of elements consumed by the SIMD fast-lane in the latter case is near zero, explaining the lack of speedup for that specific SuiteSparse matrix.

\begin{figure*}[h]
    \centering
    \includegraphics[width=1\linewidth]{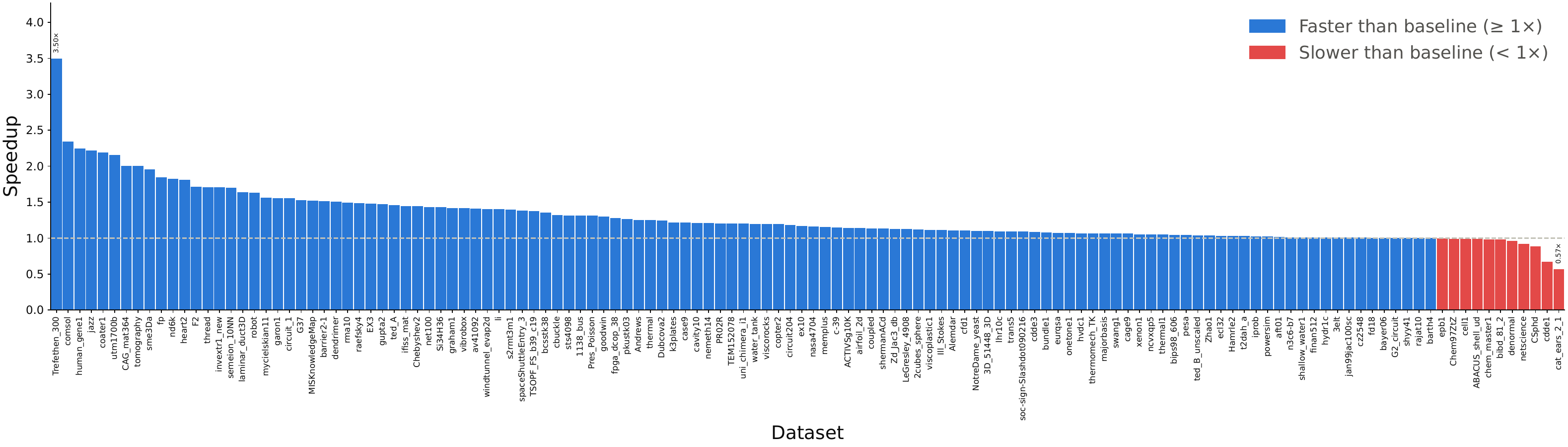}
    \caption{Speedup of Splyce for SpGEMM ($A_{ij} = \sum_{k}\textbf{B}_{ik}\cdot\textbf{C}_{kj}$).
    The same SuiteSparse matrix is used for both input matrices $\textbf{B}$ and $\textbf{C}$. 
    To represent the SuiteSparse Matrix Collection, we selected one matrix from each group, specifically choosing the one with the median non-zero density.
    }
    \label{fig:spgemm_realworld_speedup}
\end{figure*}

\begin{figure*}[h]
    \centering
    \includegraphics[width=1\linewidth]{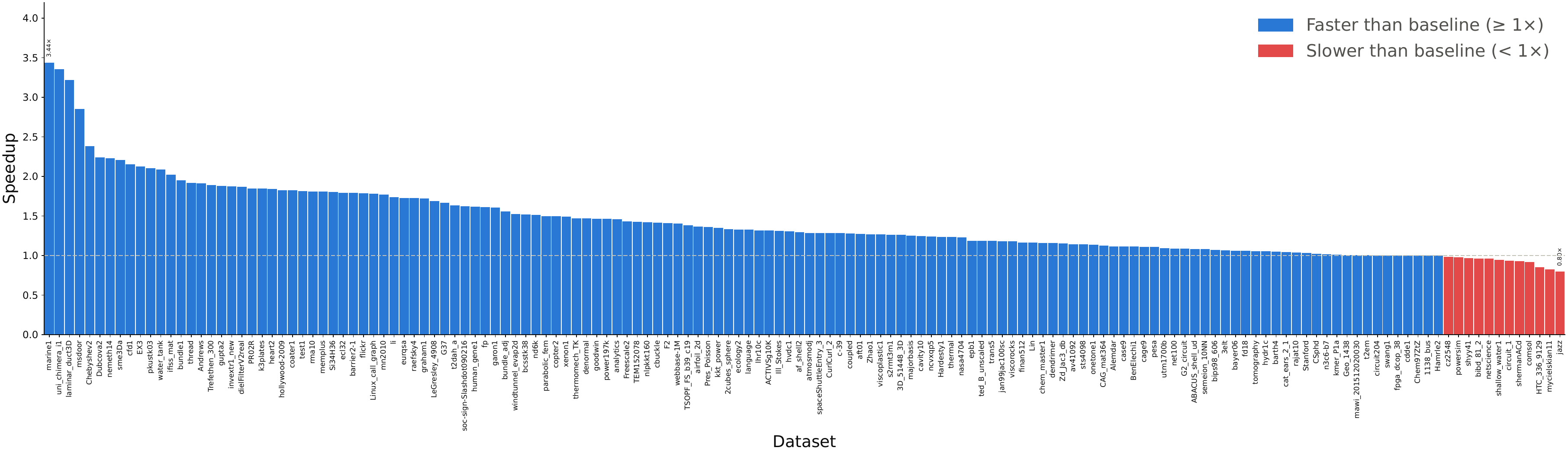}
    \caption{Speedup of Splyce for SpMSpV ($y_{i} = \sum_{j}\textbf{B}_{ij}\cdot\textbf{x}_{j}$). Input matrices follow the same selection strategy used for SpGEMM. The synthetic input vector is generated to match the sparsity factor of its corresponding matrix, subject to a minimum sparsity floor of 0.001\%.
    }
    \label{fig:spmspv_realworld_speedup}
\end{figure*}

\begin{figure*}[h]
    \centering
    \includegraphics[width=0.73\linewidth]{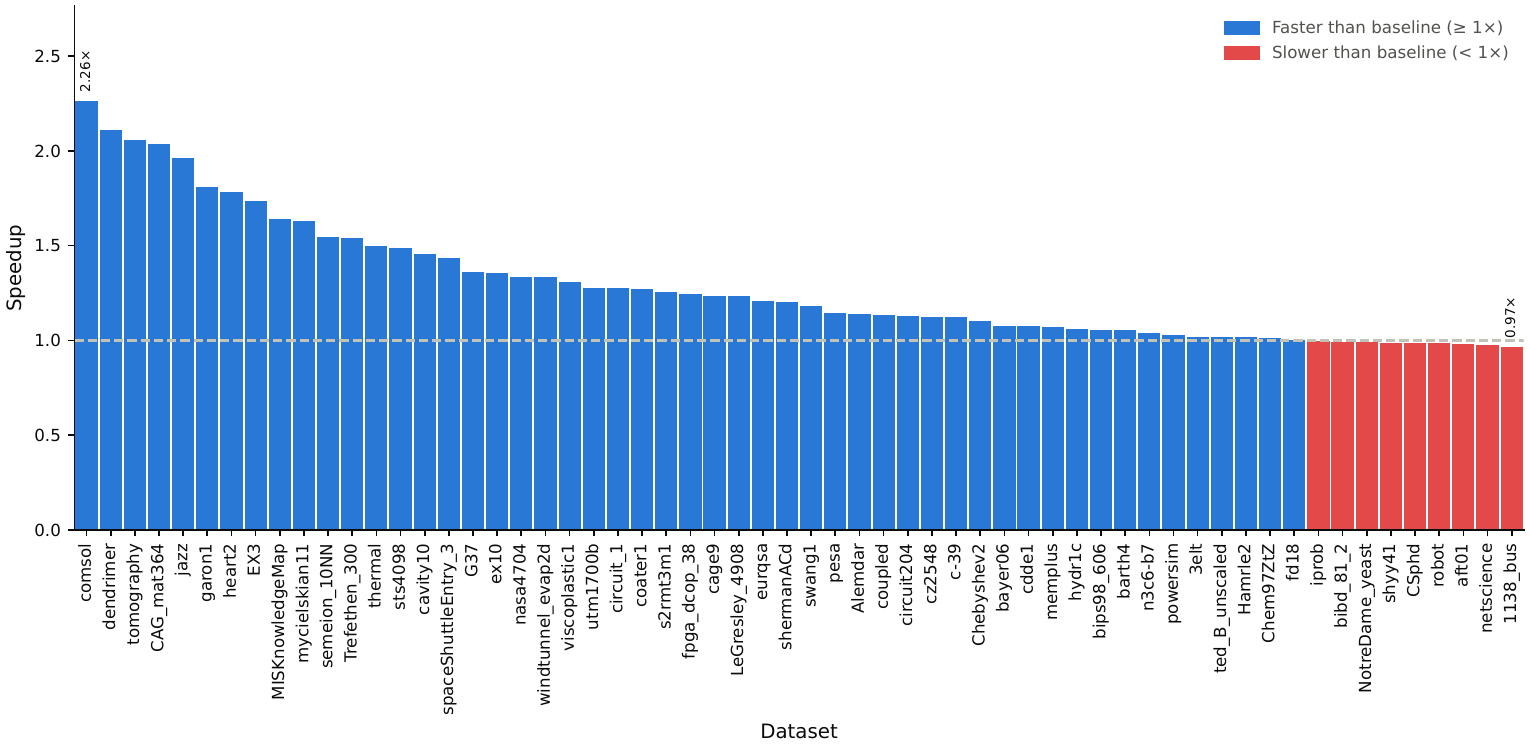}
    \caption{Speedup of Splyce for SpMTTKRP ($A_{ik} = \sum_{k,l}\textbf{B}_{ikl}\cdot\textbf{C}_{lj}\cdot\textbf{D}_{kj}$). The 2D matrices follow the same selection strategy used for SpGEMM, with the same matrix used for both inputs $\textbf{C}$ and $\textbf{D}$. The synthetic 3D input tensor is generated to match the sparsity factor of these matrices, subject to a minimum sparsity floor of 0.001\%.
    }
    \label{fig:spmttkrp_realworld_speedup}
\end{figure*}

At the same time, the \textit{SIMD(\%)} metric alone is not a definitive performance predictor, since SIMD lanes may still process structural zero values within their wide execution width.
Indeed, our extended evaluation across SpGEMM (Figure~\ref{fig:spgemm_realworld_speedup}), SpMSpV (Figure~\ref{fig:spmspv_realworld_speedup}), and SpMTTKRP (Figure~\ref{fig:spmttkrp_realworld_speedup}) kernels reveals that while \name delivers significant speedups across the majority of diverse SuiteSparse matrices, a small subset incurs a negative speedup.
Furthermore, this behaviour is highly kernel-dependent: a dataset that trigger a slowdown in one operation may still accelerate in another.
For example, the \texttt{cdde1} dataset exhibits a performance regression during the SpGEMM kernel, yet achieves positive speedup in both SpMSpV and SpMTTKRP.
Because we currently lack a definitive metric to capture these complex dataset-kernel interactions, identifying a priori which data properties trigger regression for specific sparse computations remains an open research problem.
Nevertheless, by coordinating these paths, \name broadly preserves the performance benefits harvested in dense regions without suffering significant degradation in the vast majority of sparse environments.
Because execution seamlessly falls through to the epilogue when the vector-width check (\texttt{condition\_c}) is not met, the system avoids the speculative overhead and register-packing penalties that would otherwise occur where vectorization is not viable.

\begin{table}[t]
\caption{Performance of \name on real-world sparse tensor datasets. Bold inputs are real-world matrix and the rest are synthetic with matching sparsity with a minimum of 0.001\%.}
\label{tab:real_tensor_results}
\centering

\footnotesize
\setlength{\tabcolsep}{1pt}
\renewcommand{\arraystretch}{0.9}

\begin{tabular}{
    @{} l l l r r r r
    @{}
}
\toprule
\textbf{Kernel} &
\textbf{Dataset} &
\textbf{Sparsity(\%)} &
\textbf{Base.(s)} & 
\textbf{Optim.(s)} & 
\textbf{Speedup} &
\textbf{SIMD(\%)} \\
\midrule

\multirow{4}{*}{SpGEMM}
 & internet (\textbf{B,C}) & 1.33e-3 & 210.12 & 210.09 & 1.00$\times$ & 3.45 \\
 & exdata\_1 (\textbf{B,C}) & 3.16 & 107.77 & 99.61 & 1.08$\times$ & 24.85 \\
 & c8\_mat11 (\textbf{B}) & 9.37 & 115.47 & 51.79 & 2.23$\times$ & 97.90 \\
 & heart1 (\textbf{B,C}) & 10.97 & 33.86 & 17.73 & 1.91$\times$ & 98.01 \\
\midrule

\multirow{4}{*}{SpMSpV}
 & stokes (\textbf{B}) & 2.66e-4 & 7.94 & 4.07 & 1.71$\times$ & 67.77 \\
 & mycielskian18 (\textbf{B}) & 0.78 & 2.79 & 1.66 & 1.68$\times$ & 76.63 \\
 & arabic-2005 (\textbf{B}) & 1.24e-4 & 14.87 & 8.84 & 1.68$\times$ & 81.37 \\
 & hugetrace-00020 (\textbf{B}) & 1.87e-5 & 7.07 & 6.84 & 1.03$\times$ & 0.00 \\

\midrule

\multirow{4}{*}{SpMTTKRP}
 & heart1 (\textbf{D}) & 10.97 & 1385.74 & 636.46 & 2.18$\times$ & 98.57 \\
 & CAG\_mat364 (\textbf{C,D}) & 10.25 & 5.06 & 2.46 & 2.06$\times$ & 87.13 \\
 & struct4 (\textbf{C,D}) & 1.26 & 78.28 & 49.51 & 1.58$\times$ & 95.86 \\
 & cavity26 (\textbf{C,D}) & 0.66 & 20.26 & 14.48 & 1.40$\times$ & 92.03 \\

\midrule

\multirow{4}{*}{SpMMH}
 & bayer01 (\textbf{B,D}) & 8.33e-3 & 51.84 & 54.60 & 0.95$\times$ & 21.75 \\
 & msc23052 (\textbf{B,D}) & 0.22 & 193.47 & 114.05 & 1.70$\times$ & 98.30 \\
 & smt (\textbf{B,D}) & 0.57 & 526.06 & 438.67 & 1.20$\times$ & 99.38 \\
 & mark3jac020 (\textbf{B,D}) & 6.74e-2 & 1.98 & 1.90 & 1.04$\times$ & 41.03 \\

\midrule

\multirow{4}{*}{SpTTSpM}
 & barth4 (\textbf{C}) & 6.48e-2 & 25.07 & 24.14 & 1.04$\times$ & 45.71 \\
 & rdist1 (\textbf{C}) & 0.55 & 129.13 & 86.21 & 1.50$\times$ & 84.43 \\
 & psmigr\_1 (\textbf{C}) & 0.51 & 1256.14 & 562.02 & 2.24$\times$ & 95.25 \\
 & EX6 (\textbf{C}) & 0.69 & 620.91 & 288.39 & 2.15$\times$ & 95.46 \\

\bottomrule
\end{tabular}
\vspace{-1.5em}
\end{table}

\subsection{Parallel Scalability} \label{sec:parallel_scale}

To investigate the interaction between our vectorization pass and TLP, we evaluated the scalability of the proposed transformation on a 32-core NUMA node. 
For this analysis, we selected the SpMTTKRP kernel using the same data and configurations as defined in Table~\ref{tab:real_tensor_results}. 
This kernel was chosen as a representative stress test because its computational intensity provides a measurable throughput footprint even at high core counts, whereas lower-order kernels often collapse into microsecond execution ranges that are susceptible to measurement noise.

We parallelized the outermost \texttt{scf.parallel} loop while maintaining the \textsc{Splyce}-optimized inner coiteration loop. 
As illustrated in Figure~\ref{fig:mttkrp_multicore}, the results demonstrate that \name is inherently orthogonal to TLP, with synthetic data. 
The implementation achieves near-linear self-scaling, reaching a 31.18$\times$ speedup on 32 cores relative to the single-core \name execution. 
More importantly, the absolute speedup over the original unoptimized baseline scales proportionally, reaching 67.90$\times$ on 32 cores. 
This confirms that the branchless, dual-path architecture of \name operates strictly within the instruction-level parallelism (ILP) domain. 
By resolving the structural conflicts of sparse traversal at the register level, \name saturates execution units without introducing resource contention or synchronization overhead, thereby preserving macro-scale multicore efficiency.
However, it should be noted that this near-linear scalability relies on the uniform workload distribution of synthetic data. 
Real-world sparse matrices typically exhibit irregular non-zero distributions across rows, which inherently introduces thread-level load imbalance and sub-linear scaling in practice.

\input{figure/mttkrp_core_speedup}


\section{Related Work} \label{sec:related_work}

Early efforts in sparse code generation~\cite{bik1993compilation, saddaySpmv, Karypis1997} evolved into the formalized merge lattice theory of TACO, which rivaled library-based approaches. 
This was subsequently expanded by the level-format abstraction, which modularized tensor dimensions into per-level types (\eg dense, compressed). 
By decoupling storage representation from the iteration engine, this abstraction enabled unified code generation across disparate layouts like CSR, CSF, and COO. 
Contemporary frameworks---including Finch~\cite{ahrens2025finch}, SparseTIR~\cite{ye2023sparsetir}, and the MLIR \texttt{SparseTensor}~\cite{mlir_sparse} dialect—build upon this modular blueprint to facilitate deeper integration with general-purpose ecosystems.

While vectorization is a standard optimization~\cite{xie2018cvr, sparse2022lcm_vec, spmmtc, marcos2023pact}, existing strategies predominantly accelerate dense-sparse interactions (e.g., SpMV). 
These methods typically enforce regular access patterns through blocking or explicit memory-level zero-padding, which compromises compression benefits and inflates memory bandwidth requirements. 
While hand-tuned libraries (Intel MKL, cuSPARSE) and specialized hardware accelerators deliver peak performance, they inherently sacrifice the flexibility and format-agnostic composability provided by a generalized compiler infrastructure.

Traditional auto-vectorizers successfully handle interleaved data with constant strides, but conservatively assume branches are divergent. 
To address this, VecRC~\cite{vecRC} introduces an auto-vectorizer that uses run-time checks to identify dynamic uniformity---cases where all SIMD lanes follow the same control path at run-time. 
VecRC leverages this uniformity to relax constraints on loops with control-dependent loop-carried dependencies, enabling vectorization without the overhead of speculation. 
While VecRC’s probability-based cost model effectively mitigates predication overhead in uniform scenarios, standard auto-vectorizers remain structurally incapable of handling the dynamic, data-dependent intersections inherent in sparse \textit{two-finger} loops.

The problem of two-finger loops is a well-known algorithmic bottleneck across various domains, including data compression \cite{lemire2015decoding}, relational databases \cite{graefe1993query, boncz2005monetdb}, and sequence alignment in bioinformatics \cite{smith1981identification}. 
While extensive SIMD vectorization efforts have successfully accelerated these co-iteration patterns in compression and database sort-merge joins, these optimizations do not trivially translate to sparse tensor compilers. 
Unlike standard applications that traverse contiguous arrays to materialize matching identifiers, sparse tensor coiteration navigates coordinate arrays---imposing heavy memory indirection---and tightly fuses this unpredictable intersection logic directly with floating-point arithmetic. 
This limits the applicability of traditional SIMD techniques and necessitates the specialized vectorization pass proposed in this work.

\section{Discussion and Future Work} \label{sec:discussion}

Although \name{} supports compressed outputs, current evaluation materializes dense outputs to reduce measurement overhead. 
The current transformation targets two-way intersection coiteration while the union-based one is left for future work. 

Multiway coiteration presents another extension of the dual-path approach. When the underlying operation permits associative decomposition, a multiway coiteration can be expressed as a sequence of two-way coiterations. This introduces additional choices in decomposition order and execution strategy. It may also change the phase-level trade-offs observed in Section~\ref{sec:phase_ablation}. Multiway coiteration can increase the amount of computation performed per traversal step, potentially amortizing the overheads associated with value extraction, packing, and reduction. Consequently, more extensive vectorization may become beneficial for such workloads.

Blocked-sparse formats are particularly relevant because they introduce fixed-size dense regions within a sparse contraction. 
These dense sub-blocks expose regular computation that can be mapped to SIMD execution while retaining sparse contraction at the block level. 
This may increase the amount of useful computation performed within each fast-lane iteration and improve the amortization of coiteration overheads.

Finally, the current implementation targets CPU SIMD execution. 
However, the underlying idea of converting irregular, branch-dependent sparse coiteration into a more regular, parallel execution pattern is not inherently CPU-specific. 
This suggests an opportunity to adapt \name{} to heterogeneous accelerators, where similar speculative and predicated execution strategies could be used to expose more parallelism and improve hardware utilization. 
\section{Conclusion} \label{sec:conclusion}

In this paper, we presented \name, an MLIR optimization pass designed to resolve the structural conflict between compressed sparse storage and modern parallel hardware. 
By transforming the divergent scalar coiteration loop into a branchless, dual-path execution model, we enable modern superscalar engines to maximize ILP and saturate parallel execution units previously starved by sequential dependencies.




\bibliographystyle{ACM-Reference-Format}
\bibliography{ref}

@article{mlir_sparse,
author = {Bik, Aart and Koanantakool, Penporn and Shpeisman, Tatiana and Vasilache, Nicolas and Zheng, Bixia and Kjolstad, Fredrik},
title = {Compiler Support for Sparse Tensor Computations in MLIR},
year = {2022},
issue_date = {December 2022},
publisher = {Association for Computing Machinery},
address = {New York, NY, USA},
volume = {19},
number = {4},
issn = {1544-3566},
url = {https://doi.org/10.1145/3544559},
doi = {10.1145/3544559},
month = sep,
articleno = {50},
numpages = {25}
}

@INPROCEEDINGS{tmayasin2014,
  author={Yasin, Ahmad},
  booktitle={2014 IEEE International Symposium on Performance Analysis of Systems and Software (ISPASS)}, 
  title={A Top-Down method for performance analysis and counters architecture}, 
  year={2014},
  volume={},
  number={},
  pages={35-44},
  doi={10.1109/ISPASS.2014.6844459}}

@article{taco,
author = {Kjolstad, Fredrik and Kamil, Shoaib and Chou, Stephen and Lugato, David and Amarasinghe, Saman},
title = {The tensor algebra compiler},
year = {2017},
issue_date = {October 2017},
publisher = {Association for Computing Machinery},
address = {New York, NY, USA},
volume = {1},
number = {OOPSLA},
url = {https://doi.org/10.1145/3133901},
doi = {10.1145/3133901},
journal = {Proc. ACM Program. Lang.},
month = oct,
articleno = {77},
numpages = {29}
}

@article{lemire2015decoding,
author = {Lemire, D. and Boytsov, L.},
title = {Decoding billions of integers per second through vectorization},
year = {2015},
issue_date = {January 2015},
publisher = {John Wiley \& Sons, Inc.},
address = {USA},
volume = {45},
number = {1},
issn = {0038-0644},
url = {https://doi.org/10.1002/spe.2203},
doi = {10.1002/spe.2203},
journal = {Softw. Pract. Exper.},
month = jan,
pages = {1–29},
numpages = {29}
}

@article{graefe1993query,
  title={Query evaluation techniques for large databases},
  author={Graefe, Goetz},
  journal={ACM Computing Surveys (CSUR)},
  volume={25},
  number={2},
  pages={73--169},
  year={1993},
  publisher={ACM New York, NY, USA}
}

@inproceedings{boncz2005monetdb,
  title={MonetDB/X100: Hyper-Pipelining Query Execution.},
  author={Boncz, Peter and Zukowski, Marcin and Nes, Niels},
  booktitle={Cidr},
  volume={5},
  pages={225--237},
  year={2005}
}

@article{smith1981identification,
  title={Identification of common molecular subsequences},
  author={Smith, Temple F and Waterman, Michael S and others},
  journal={Journal of molecular biology},
  volume={147},
  number={1},
  pages={195--197},
  year={1981},
  publisher={Elsevier Science}
}

@article{dastac,
author = {Ghorbani, Mahdi and Bauer, Emilien and Grosser, Tobias and Shaikhha, Amir},
title = {Compressed and Parallelized Structured Tensor Algebra},
year = {2025},
issue_date = {April 2025},
publisher = {Association for Computing Machinery},
address = {New York, NY, USA},
volume = {9},
number = {OOPSLA1},
url = {https://doi.org/10.1145/3720506},
doi = {10.1145/3720506},
journal = {Proc. ACM Program. Lang.},
month = apr,
articleno = {141},
numpages = {29}
}

@article{structured,
author = {Ghorbani, Mahdi and Huot, Mathieu and Hashemian, Shideh and Shaikhha, Amir},
title = {Compiling Structured Tensor Algebra},
year = {2023},
issue_date = {October 2023},
publisher = {Association for Computing Machinery},
address = {New York, NY, USA},
volume = {7},
number = {OOPSLA2},
url = {https://doi.org/10.1145/3622804},
doi = {10.1145/3622804},
month = oct,
articleno = {229},
numpages = {30}
}

@article{Senanayake2020A,
title={A sparse iteration space transformation framework for sparse tensor algebra},
author={Ryan Senanayake and Changwan Hong and Ziheng Wang and Amalee Wilson and Stephen Chou and Shoaib Kamil and Saman P. Amarasinghe and Fredrik Kjolstad},
journal={Proceedings of the ACM on Programming Languages},
year={2020},
volume={4},
pages={1 - 30},
doi={10.1145/3428226}
}

@article{Li2019Analytical,
title={Analytical Cache Modeling and Tilesize Optimization for Tensor Contractions},
author={Rui Li and Aravind Sukumaran-Rajam and R. Veras and Tze Meng Low and F. Rastello and A. Rountev and P. Sadayappan},
journal={SC19: International Conference for High Performance Computing, Networking, Storage and Analysis},
year={2019},
pages={1-13},
doi={10.1145/3295500.3356218}
}

@article{distal,
title={DISTAL: the distributed tensor algebra compiler},
author={Rohan Yadav and A. Aiken and Fredrik Kjolstad},
journal={Proceedings of the 43rd ACM SIGPLAN International Conference on Programming Language Design and Implementation},
year={2022},
doi={10.1145/3519939.3523437}
}

@article{Chen2018Learning,
title={Learning to Optimize Tensor Programs},
author={Tianqi Chen and Lianmin Zheng and Eddie Q. Yan and Ziheng Jiang and T. Moreau and L. Ceze and Carlos Guestrin and A. Krishnamurthy},
year={2018},
pages={3393-3404},
doi={}
}

@inproceedings{spmmtc,
author = {He, Xianhao and Wang, Haotian and Zhang, Jiapeng and Yang, Wangdong and Chronopoulos, Anthony Theodore and Li, Kenli},
title = {An Input-Aware Sparse Tensor Compiler Empowered by Vectorized Acceleration},
year = {2025},
isbn = {9798331503048},
publisher = {IEEE Press},
url = {https://doi.org/10.1109/DAC63849.2025.11133371},
doi = {10.1109/DAC63849.2025.11133371},
booktitle = {Proceedings of the 62nd Annual ACM/IEEE Design Automation Conference},
articleno = {437},
numpages = {7},
location = {San Francisco, California, United States},
series = {DAC '25}
}

@article{Shi2016Tensor,
title={Tensor Contractions with Extended BLAS Kernels on CPU and GPU},
author={Yang Shi and U. Niranjan and Anima Anandkumar and C. Cecka},
journal={2016 IEEE 23rd International Conference on High Performance Computing (HiPC)},
year={2016},
pages={193-202},
doi={10.1109/hipc.2016.031}
}

@inproceedings{mlircgo21,
author = {Lattner, Chris and Amini, Mehdi and Bondhugula, Uday and Cohen, Albert and Davis, Andy and Pienaar, Jacques and Riddle, River and Shpeisman, Tatiana and Vasilache, Nicolas and Zinenko, Oleksandr},
title = {MLIR: scaling compiler infrastructure for domain specific computation},
year = {2021},
isbn = {9781728186139},
publisher = {IEEE Press},
url = {https://doi.org/10.1109/CGO51591.2021.9370308},
doi = {10.1109/CGO51591.2021.9370308},
booktitle = {Proceedings of the 2021 IEEE/ACM International Symposium on Code Generation and Optimization},
pages = {2–14},
numpages = {13},
location = {Virtual Event, Republic of Korea},
series = {CGO '21}
}

@misc{intel_simd,
	title = {{SIMD vectorization in LLVM and GCC for Intel{\ifmmode\circledR\else\textregistered\fi} CPUs and GPUs}},
	journal = {Intel},
	year = {2026},
	month = apr,
	note = {[Online; accessed 17. Apr. 2026]},
	url = {https://www.intel.com/content/www/us/en/developer/articles/technical/vectorization-llvm-gcc-cpus-gpus.html}
}

@misc{fog_microarchitecture,
  author       = {Fog, Agner},
  title        = {The Microarchitecture of Intel, AMD, and VIA CPUs},
  year         = {2023},
  howpublished = {\url{https://www.agner.org/optimize/microarchitecture.pdf}}
}

@techreport{ilp_multipro,
author = {Stankovic, J. A. and Ramamritham, K. and Shiah, P. and Zhao, W.},
title = {Real-Time Scheduling Algorithms for Multiprocessors},
year = {1989},
publisher = {University of Massachusetts},
address = {USA},
}

@article{suitesparse1, doi = {10.21105/joss.01244}, url = {https://doi.org/10.21105/joss.01244}, year = {2019}, publisher = {The Open Journal}, volume = {4}, number = {35}, pages = {1244}, author = {Kolodziej, Scott P. and Aznaveh, Mohsen and Bullock, Matthew and David, Jarrett and Davis, Timothy A. and Henderson, Matthew and Hu, Yifan and Sandstrom, Read}, title = {The SuiteSparse Matrix Collection Website Interface}, journal = {Journal of Open Source Software} }

@article{suitesparse2,
author = {Davis, Timothy A. and Hu, Yifan},
title = {The university of Florida sparse matrix collection},
year = {2011},
issue_date = {November 2011},
publisher = {Association for Computing Machinery},
address = {New York, NY, USA},
volume = {38},
number = {1},
issn = {0098-3500},
url = {https://doi.org/10.1145/2049662.2049663},
doi = {10.1145/2049662.2049663},
journal = {ACM Trans. Math. Softw.},
month = dec,
articleno = {1},
numpages = {25}
}

@inproceedings{wall1991limits,
author = {Wall, David W.},
title = {Limits of instruction-level parallelism},
year = {1991},
isbn = {0897913809},
publisher = {Association for Computing Machinery},
address = {New York, NY, USA},
url = {https://doi.org/10.1145/106972.106991},
doi = {10.1145/106972.106991},
booktitle = {Proceedings of the Fourth International Conference on Architectural Support for Programming Languages and Operating Systems},
pages = {176–188},
numpages = {13},
location = {Santa Clara, California, USA},
series = {ASPLOS IV}
}

@inproceedings{marcos2023pact,
author = {Horro, Marcos and Pouchet, Louis-No\"{e}l and Rodr\'{\i}guez, Gabriel and Touri\~{n}o, Juan},
title = {Custom High-Performance Vector Code Generation for Data-Specific Sparse Computations},
year = {2023},
isbn = {9781450398688},
publisher = {Association for Computing Machinery},
address = {New York, NY, USA},
url = {https://doi.org/10.1145/3559009.3569668},
doi = {10.1145/3559009.3569668},
booktitle = {Proceedings of the International Conference on Parallel Architectures and Compilation Techniques},
pages = {160–171},
numpages = {12},
location = {Chicago, Illinois},
series = {PACT '22}
}

@inproceedings{vecRC,
author = {Liu, Bangtian and Laird, Avery and Tsang, Wai Hung and Mahjour, Bardia and Dehnavi, Maryam Mehri},
title = {Combining Run-Time Checks and Compile-Time Analysis to Improve Control Flow Auto-Vectorization},
year = {2023},
isbn = {9781450398688},
publisher = {Association for Computing Machinery},
address = {New York, NY, USA},
url = {https://doi.org/10.1145/3559009.3569663},
doi = {10.1145/3559009.3569663},
booktitle = {Proceedings of the International Conference on Parallel Architectures and Compilation Techniques},
pages = {439–450},
numpages = {12},
location = {Chicago, Illinois},
series = {PACT '22}
}

@misc{intel_avx512,
	title = {{Intel{\ifmmode\circledR\else\textregistered\fi} Xeon{\ifmmode\circledR\else\textregistered\fi} Processor Scalable Family Technical Overview}},
	journal = {Intel},
	year = {2026},
	month = may,
	note = {[Online; accessed 1. May 2026]},
	url = {https://www.intel.com/content/www/us/en/developer/articles/technical/xeon-processor-scalable-family-technical-overview.html}
}

@article{seznec_branchpred,
  TITLE = {{A case for (partially) tagged geometric history length branch prediction}},
  AUTHOR = {Seznec, Andr{\'e} and Michaud, Pierre},
  URL = {https://inria.hal.science/hal-03408381},
  JOURNAL = {{The Journal of Instruction-Level Parallelism}},
  PUBLISHER = {{North Carolina State University}},
  VOLUME = {8},
  PAGES = {23},
  YEAR = {2006},
  MONTH = Feb,
  HAL_ID = {hal-03408381},
  HAL_VERSION = {v1},
}

@book{comp_ar,
author = {Hennessy, John L. and Patterson, David A.},
title = {Computer Architecture, Fifth Edition: A Quantitative Approach},
year = {2011},
isbn = {012383872X},
publisher = {Morgan Kaufmann Publishers Inc.},
address = {San Francisco, CA, USA},
edition = {5th}
}

@article{zhao2022polyhedral,
  title={Polyhedral specification and code generation of sparse tensor contraction with co-iteration},
  author={Zhao, Tuowen and Popoola, Tobi and Hall, Mary and Olschanowsky, Catherine and Strout, Michelle},
  journal={ACM Transactions on Architecture and Code Optimization},
  volume={20},
  number={1},
  pages={1--26},
  year={2022},
  publisher={ACM New York, NY}
}

@ARTICLE{hardware_accel_sparse,
  author={Dave, Shail and Baghdadi, Riyadh and Nowatzki, Tony and Avancha, Sasikanth and Shrivastava, Aviral and Li, Baoxin},
  journal={Proceedings of the IEEE}, 
  title={Hardware Acceleration of Sparse and Irregular Tensor Computations of ML Models: A Survey and Insights}, 
  year={2021},
  volume={109},
  number={10},
  pages={1706-1752},
  doi={10.1109/JPROC.2021.3098483}
}

@article{Kuhn2023Sep,
	author = {K{\ifmmode\ddot{u}\else\"{u}\fi}hn, Martin J. and Holke, Johannes and Lutz, Annette and Thies, Jonas and R{\ifmmode\ddot{o}\else\"{o}\fi}hrig-Z{\ifmmode\ddot{o}\else\"{o}\fi}llner, Melven and Bleh, Alexander and Backhaus, Jan and Basermann, Achim},
	title = {{SIMD vectorization for simultaneous solution of locally varying linear systems with multiple right-hand sides}},
	journal = {J. Supercomput.},
	volume = {79},
	number = {13},
	pages = {14684--14706},
	year = {2023},
	month = sep,
	issn = {1573-0484},
	publisher = {Springer US},
	doi = {10.1007/s11227-023-05220-4}
}

@misc{intelMKL,
	title = {{Accelerate Fast Math with Intel{\ifmmode\circledR\else\textregistered\fi} oneAPI Math Kernel Library}},
	journal = {Intel},
	year = {2026},
	month = apr,
	note = {[Online; accessed 16. Apr. 2026]},
	url = {https://www.intel.com/content/www/us/en/developer/tools/oneapi/onemkl.html}
}

@misc{cusparse,
	title = {{1. Introduction {\ifmmode---\else\textemdash\fi} cuSPARSE 13.2 documentation}},
	year = {2026},
	month = apr,
	note = {[Online; accessed 16. Apr. 2026]},
	url = {https://docs.nvidia.com/cuda/cusparse/index.html}
}

@article{kolda2009tensor,
author = {Kolda, Tamara G. and Bader, Brett W.},
title = {Tensor Decompositions and Applications},
journal = {SIAM Review},
volume = {51},
number = {3},
pages = {455-500},
year = {2009},
doi = {10.1137/07070111X},
URL = {https://doi.org/10.1137/07070111X},
eprint = {https://doi.org/10.1137/07070111X}
}

@article{hirata2003tensor,
author = {Hirata, So},
title = {Tensor Contraction Engine: Abstraction and Automated Parallel Implementation of Configuration-Interaction, Coupled-Cluster, and Many-Body Perturbation Theories},
journal = {The Journal of Physical Chemistry A},
volume = {107},
number = {46},
pages = {9887-9897},
year = {2003},
doi = {10.1021/jp034596z},
URL = {https://doi.org/10.1021/jp034596z},
eprint = {https://doi.org/10.1021/jp034596z}
}

@article{williams2009roofline,
author = {Williams, Samuel and Waterman, Andrew and Patterson, David},
title = {Roofline: an insightful visual performance model for multicore architectures},
year = {2009},
issue_date = {April 2009},
publisher = {Association for Computing Machinery},
address = {New York, NY, USA},
volume = {52},
number = {4},
issn = {0001-0782},
url = {https://doi.org/10.1145/1498765.1498785},
doi = {10.1145/1498765.1498785},
journal = {Commun. ACM},
month = apr,
pages = {65–76},
numpages = {12}
}

@inproceedings{nuzman2006auto,
author = {Nuzman, Dorit and Rosen, Ira and Zaks, Ayal},
title = {Auto-vectorization of interleaved data for SIMD},
year = {2006},
isbn = {1595933204},
publisher = {Association for Computing Machinery},
address = {New York, NY, USA},
url = {https://doi.org/10.1145/1133981.1133997},
doi = {10.1145/1133981.1133997},
booktitle = {Proceedings of the 27th ACM SIGPLAN Conference on Programming Language Design and Implementation},
pages = {132–143},
numpages = {12},
location = {Ottawa, Ontario, Canada},
series = {PLDI '06}
}

@article{Qin2021Extending,
title={Extending Sparse Tensor Accelerators to Support Multiple Compression Formats},
author={Eric Qin and Geonhwa Jeong and William Won and Sheng-Chun Kao and Hyoukjun Kwon and S. Srinivasan and Dipankar Das and G. Moon and S. Rajamanickam and T. Krishna},
journal={2021 IEEE International Parallel and Distributed Processing Symposium (IPDPS)},
year={2021},
pages={1014-1024},
doi={10.1109/ipdps49936.2021.00110}
}

@article{dongarra1990set,
author = {Dongarra, J. J. and Du Croz, Jeremy and Hammarling, Sven and Duff, I. S.},
title = {A set of level 3 basic linear algebra subprograms},
year = {1990},
issue_date = {March 1990},
publisher = {Association for Computing Machinery},
address = {New York, NY, USA},
volume = {16},
number = {1},
issn = {0098-3500},
url = {https://doi.org/10.1145/77626.79170},
doi = {10.1145/77626.79170},
journal = {ACM Trans. Math. Softw.},
month = mar,
pages = {1–17},
numpages = {17}
}

@article{Chen2019Performance-Aware,
title={Performance-Aware Model for Sparse Matrix-Matrix Multiplication on the Sunway TaihuLight Supercomputer},
author={Yuedan Chen and Kenli Li and Wangdong Yang and Guoqing Xiao and Xianghui Xie and Tao Li},
journal={IEEE Transactions on Parallel and Distributed Systems},
year={2019},
volume={30},
pages={923-938},
doi={10.1109/tpds.2018.2871189}
}

@article{Chou2018Format,
title={Format abstraction for sparse tensor algebra compilers},
author={Stephen Chou and Fredrik Kjolstad and Saman P. Amarasinghe},
journal={Proceedings of the ACM on Programming Languages},
year={2018},
volume={2},
pages={1 - 30},
doi={10.1145/3276493}
}

@article{Smith2015Tensor-matrix,
title={Tensor-matrix products with a compressed sparse tensor},
author={Shaden Smith and G. Karypis},
journal={Proceedings of the 5th Workshop on Irregular Applications: Architectures and Algorithms},
year={2015},
doi={10.1145/2833179.2833183}
}

@article{Willcock2006Accelerating,
title={Accelerating sparse matrix computations via data compression},
author={Jeremiah Willcock and A. Lumsdaine},
year={2006},
pages={307-316},
doi={10.1145/1183401.1183444}
}

@inproceedings{bik1993compilation,
  title={Compilation techniques for sparse matrix computations},
  author={Bik, Aart JC and Wijshoff, Harry AG},
  booktitle={Proceedings of the 7th international conference on Supercomputing},
  pages={416--424},
  year={1993}
}

@misc{Karypis1997,
	author = {Karypis, George and Kumar, Vipin},
	title = {{METIS: A Software Package for Partitioning Unstructured Graphs, Partitioning Meshes, and Computing Fill-Reducing Orderings of Sparse Matrices}},
	year = {1997},
	note = {[Online; accessed 28. Apr. 2026]},
	url = {https://conservancy.umn.edu/items/2f610239-590c-45c0-bcd6-321036aaad56}
}

@INPROCEEDINGS{saddaySpmv,
  author={White, J.B. and Sadayappan, P.},
  booktitle={Proceedings Fourth International Conference on High-Performance Computing}, 
  title={On improving the performance of sparse matrix-vector multiplication}, 
  year={1997},
  volume={},
  number={},
  pages={66-71},
  doi={10.1109/HIPC.1997.634472}
}

@article{ahrens2025finch,
  title={Finch: Sparse and structured tensor programming with control flow},
  author={Ahrens, Willow and Collin, Teodoro Fields and Patel, Radha and Deeds, Kyle and Hong, Changwan and Amarasinghe, Saman},
  journal={Proceedings of the ACM on Programming Languages},
  volume={9},
  number={OOPSLA1},
  pages={1042--1072},
  year={2025},
  publisher={ACM New York, NY, USA}
}

@inproceedings{ye2023sparsetir,
  title={Sparsetir: Composable abstractions for sparse compilation in deep learning},
  author={Ye, Zihao and Lai, Ruihang and Shao, Junru and Chen, Tianqi and Ceze, Luis},
  booktitle={Proceedings of the 28th ACM International Conference on Architectural Support for Programming Languages and Operating Systems, Volume 3},
  pages={660--678},
  year={2023}
}

@inproceedings{xie2018cvr,
  title={Cvr: Efficient vectorization of spmv on x86 processors},
  author={Xie, Biwei and Zhan, Jianfeng and Liu, Xu and Gao, Wanling and Jia, Zhen and He, Xiwen and Zhang, Lixin},
  booktitle={Proceedings of the 2018 International Symposium on Code Generation and Optimization},
  pages={149--162},
  year={2018}
}

@inproceedings{sparse2022lcm_vec,
author = {Cheshmi, Kazem and Cetinic, Zachary and Dehnavi, Maryam Mehri},
title = {Vectorizing sparse matrix computations with partially-strided codelets},
year = {2022},
isbn = {9784665454445},
publisher = {IEEE Press},
booktitle = {Proceedings of the International Conference on High Performance Computing, Networking, Storage and Analysis},
articleno = {32},
numpages = {15},
location = {Dallas, Texas},
series = {SC '22}
}
\appendix
\section{Artifact Appendix}

\subsection{Abstract}

This artifact contains the implementation of \name, an MLIR pass that vectorizes sparse coiteration (\texttt{scf.while} two-finger loops) by transforming them into a dual-path version of branchless fast-lane and scalar-epilogue.
The artifact includes the \name MLIR pass (built against LLVM/MLIR mainline), the five benchmark kernels evaluated in the paper (SpGEMM, SpMSpV, SpMTTKRP, SpMMH, SpTTSpM), synthetic sparse tensor generators, scripts to download and preprocess the SuiteSparse matrices used in Table~\ref{tab:real_tensor_results}, and analysis scripts that reproduce the phase-ablation study (Table~\ref{tab:spgemm_ablation}, Figure~\ref{fig:tma_results}), the speedup results (Table~\ref{tab:basic_kernels}, Table~\ref{tab:real_tensor_results}), the sparsity scaling sweep (Figure~\ref{fig:spgemm_runtime_sparsity}), the vector width experiment (Figure~\ref{fig:spgemm_vectorsize}), and the parallel scalability study (Figure~\ref{fig:mttkrp_multicore}).
Reproducing these results validates the following central claims of the paper:
\begin{itemize}[leftmargin=*]
    \item \name reduces branch misprediction.
    \item In \name, phase config \confnum{5} is the best performing one on the particular hardware.
    \item \name achieves a 2.38$\times$ geomean speedup on synthetic data.
    \item \name sustains speedups on real-world SuiteSparse inputs.
    \item \name constrains its optimization within a single core.
\end{itemize}

Full instructions for environment configuration, compilation, and experiment reproduction are detailed in \texttt{README.md} of the repository. 
This appendix directs readers to specific sections of that online documentation for detailed execution steps.

\subsection{Artifact checklist}

{\small
\begin{itemize}
  \item {\bf Algorithm: } Dual-path (vectorized fast-lane + scalar epilogue) SIMD coiteration transform for sparse tensor contractions.
  \item {\bf Program: } \name MLIR pass (C++/MLIR), benchmark drivers for SpGEMM, SpMSpV, SpMTTKRP, SpMMH, SpTTSpM.
  \item {\bf Compilation: } LLVM/MLIR 23; \texttt{-O3} \texttt{-mavx512f} \texttt{-mavx512vl}
  \item {\bf Transformations: } \texttt{SparseTensor} $\rightarrow$ \texttt{SCF/Vector} $\rightarrow$ \texttt{LLVM-IR} intercepted by Splyce between \texttt{SCF/Vector} dialect stages.
  \item {\bf Binary: } \texttt{splyce-opt} MLIR pass plugin, benchmark executables (pre-built on the server; source also provided for inspection/rebuild).
  \item {\bf Data set: } Synthetic uniform-random sparse tensors; real-world matrices from the SuiteSparse Matrix Collection ($\eg$ \textit{internet, exdata\_1, c8\_mat11, heart1, stokes, bayer01, barth4})
  \item {\bf Run-time environment: } Ubuntu 24.04.4 LTS, Linux kernel 6.x (as configured on the provided server)
  \item {\bf Hardware: } Provided server---dual-socket Intel Xeon Gold 6430 (Sapphire Rapids), 64 physical cores (32/socket), AVX-512, 120 MB shared L3, 2 MB L2/core; SMT disabled, threads pinned to a single NUMA node.
  \item {\bf Execution: } Single-threaded for Section~\ref{sec:phase_ablation}-~\ref{sec:sparsity_scaling}; multi-threaded (1-32 cores) for Section~\ref{sec:parallel_scale}.
  \item {\bf Metrics: } Wall-clock execution time, speedup vs. the pre-built unmodified MLIR baseline, IPC, branch misprediction count, TMA pipeline-slot breakdown.
  \item {\bf Output: } CSV files with timings, plots matching Figures~\ref{fig:tma_results}-\ref{fig:mttkrp_multicore}
  \item {\bf Experiments: } Phase ablation (Table~\ref{tab:spgemm_ablation}, Figure~\ref{fig:tma_results}), vectorization overhead across 5 kernels (Table~\ref{tab:basic_kernels}), real-world dataset evaluation (Table~\ref{tab:real_tensor_results}), Vector width ablation (Figure~\ref{fig:spgemm_vectorsize}), Sparsity Scaling (Figure~\ref{fig:spgemm_runtime_sparsity}), Multicore Scaling (Figure~\ref{fig:mttkrp_multicore}).
  \item {\bf How much disk space required (approximately)?: } $\sim$116 GiB (Mostly for SuiteSparse and Synthetic matrices)
  \item {\bf Publicly available?: } Yes
  \item {\bf Code licenses (if publicly available)?: } Apache 2.0 (matching LLVM/MLIR licensing)
  \item {\bf Data licenses (if publicly available)?: } SuiteSparse matrices are publicly available under their respective collection terms.
  \item {\bf Workflow automation framework used?: } Shell Scripts, Python Scripts, CMake
  \item {\bf Archived (provide DOI)?: } \url{https://doi.org/10.5281/zenodo.21878752}
  \item {\bf Archived (Trovi): } \url{https://trovi.chameleoncloud.org/dashboard/artifacts/4ad43ccf-564d-450a-abf2-25a99d5d3d70}
  \item {\bf Repository: } \url{https://github.com/KabilanMA/Splyce}
\end{itemize}
}

\subsection{Description}

\subsubsection{Access}

\textbf{Zenodo}: \url{https://doi.org/10.5281/zenodo.21878752}

\subsubsection{Hardware dependencies}

Reproducing the absolute timings reported in this work requires an \textit{x86-64} processor with \textit{AVX-512F} and \textit{AVX-512VL} support. 
All experiments were conducted on a dual-socket Intel Xeon Gold 6430 (Sapphire Rapids) server running Ubuntu 24.04.4 LTS. 
The system features 64 total physical cores (32 per socket), 120\,MB of shared L3 cache, and 2\,MB of dedicated L2 cache per core. 
To isolate microarchitectural behavior and ensure deterministic, cycle-accurate profiling, Simultaneous Multithreading (SMT) was disabled, and all execution threads were pinned to a single NUMA node. 
Consequently, multi-socket hardware is not strictly required for reproduction. 
The \name optimization pass is implemented and evaluated against the LLVM/MLIR \texttt{23.0.0-git} mainline.

Regarding core counts, the single-thread evaluations in Sections~\ref{sec:phase_ablation} through \ref{sec:sparsity_scaling} require only a single core. 
Conversely, the scalability evaluation in Section~\ref{sec:parallel_scale} requires a dense multi-core environment. 
At least 32 physical cores on a single NUMA node are necessary to reproduce the full sweep shown in Figure~\ref{fig:mttkrp_multicore}, although the general scaling trend remains observable on hardware with fewer cores.

\subsubsection{Software dependencies}

We defer to the \texttt{README.md} within the artifact repository for complete and up-to-date specifications regarding software dependencies and environment configuration.

\subsubsection{Containerized Evaluation}

In addition to the pre-configured server, we supply a \texttt{Dockerfile} and supporting documentation to facilitate local execution. 
Reviewers electing to use this method should be advised of two caveats. 
First, building the Docker image requires compiling LLVM from source, a process that can be highly time-consuming depending on the host machine. 
Second, while this environment is fully capable of functional validation, we cannot guarantee the fidelity of the absolute execution timings or scaling metrics due to the inherent overhead of containerization.

\subsubsection{Data sets}

The evaluation utilizes both synthetic and real-world datasets. Synthetic tensors, featuring uniform random sparsity, are generated on-demand via the \texttt{experiments/gen\_data.sh} script. 
Real-world matrices are retrieved dynamically from the SuiteSparse Matrix Collection using automated scripts included in the repository.

\subsection{Installation}

Detailed procedures for resolving the LLVM dependency--either by compiling from source or configuring a pre-existing build--are provided in the \texttt{README.md} file. 
The documentation subsequently outlines the compilation steps for Splyce, along with a \textit{Getting Started} section designed to simply validate the environment setup through a minimal working example.

\subsection{Experiment workflow}

To facilitate reproducibility, the artifact provides six automated scripts capable of generating all plots and corresponding CSV data files discussed in the Evaluation section.
We also supply reference plots and baseline CSV output against which reviewers can validate their generated results.
Comprehensive documentation detailing each experiment is maintained in the \texttt{README.md} of the repository.

\subsection{Evaluation and expected results}

Every experimental workflow is isolated within a specific folder under the \texttt{experiments} directory.
These folders contain the baseline execution numbers and reference plots. 
Upon executing the provided scripts, reviewers can validate their setup by ensuring their newly generated data and figures closely match these reference files, which were used to populate the tables and figures in the paper.

\end{document}